\documentclass[useAMS,usenatbib]{mn2e}
\newcommand \kms{km s$^{-1}$}
\newcommand \zabs{z$_{\rm abs}$}
\newcommand \zem{z$_{\rm em}$}
\newcommand \url{}

\newcommand{\ion}[1]{~\textsc{#1}}

\usepackage{xcolor}

\usepackage{longtable}
\usepackage{graphicx,subcaption}
\usepackage{lscape}
\usepackage{rotating}
\usepackage{amssymb}
\usepackage{amsmath}
\usepackage{adjustbox}
\usepackage{hyperref}
\usepackage[T1]{fontenc} 
\usepackage{lmodern} 
\rmfamily 
\DeclareFontShape{T1}{lmr}{b}{sc}{<->ssub*cmr/bx/sc}{}
\DeclareFontShape{T1}{lmr}{bx}{sc}{<->ssub*cmr/bx/sc}{}

\def \HI{H\,\textsc{i}}
\def \OI{O\,\textsc{i}}
\def \cloudy{\textsc{Cloudy}}
\def \nodata{. . .}

\def \vninety{$\Delta V_{90}$}
\def \subIC{$_{\rm IC}$}
\def \IC{IC($X_{i}$)}

\def \Zcloudy{[M/H]$_{\rm cld}$}

\def \nH{$n_{\rm H}$}
\def \deltaV{${\rm \Delta V}$}
\def \Msol{M$_\odot$}

\begin{document}

\title[Chemical enrichment of metal-poor subDLAs]{Unmixed metals: Variations in the enrichment of z~$\sim4$ sub-damped Lyman $\alpha$ systems}

\author[Berg et al.] {
\parbox[t]{\textwidth}{
Trystyn A. M. Berg$^{1}$, Louise A. Welsh$^{2,3}$, Ryan J. Cooke$^{3}$, Lise Christensen$^{4}$, Valentina D'Odorico$^{2,5}$, Sara L. Ellison$^6$, Sebasti\'an L\'opez$^{7}$\\\
}
\\
$^{1}$NRC Herzberg Astronomy and Astrophysics Research Centre, 5071 West Saanich Road, Victoria, B.C., Canada, V9E 2E7\\
$^{2}$INAF-Osservatorio Astronomico di Trieste, Via Tiepolo 11, I-34143 Trieste, Italy.\\
$^{3}$Centre for Extragalactic Astronomy, Durham University, South Road, Durham, DH1 3LE, UK \\
$^{4}$Cosmic Dawn Center, Niels Bohr Institute, University of Copenhagen, Jagtvej 128, 2200-N Copenhagen, Denmark\\
$^{5}$Scuola Normale Superiore Piazza dei Cavalieri, 7 I-56126 Pisa, Italy.\\
$^{6}$Department of Physics and Astronomy, University of Victoria, Victoria, British Columbia, V8P 1A1, Canada.\\
$^{7}$Departamento de Astronom\'{\i}a, Universidad de Chile, Casilla 36-D, Santiago, Chile.\\
}

\maketitle
\begin{abstract}
The chemical abundance patterns of near-pristine objects provide important constraints on the properties of the first generations of stars in the Universe. We present the chemical abundances of five very metal-poor ([M/H]$<-2.5$) sub damped Ly$\alpha$ systems (subDLAs) covering the redshift range $3.6<z<4.3$, identified with the XQ-100 survey. We find that the subDLAs in our sample show consistent chemical abundance patterns (in particular [C/O], [Al/O], and [Fe/O]) with those of very metal-poor DLAs.  Based on Voigt profile fitting, the chemical abundance ratios [C/O], [Al/O], and [Si/O] of individual velocity components in at least three of the subDLAs shows some intrinsic scatter. In order to verify these chemical inhomogeneities in absorption components, we present a novel method for computing ionization corrections (ICs) on a component-by-component basis and show that ICs alone cannot explain the variations in [C/O], [Al/O], and [Si/O] between components of the same absorber at $\approx2\sigma$ significance. Comparing the observed abundance ratios to the simulated core-collapse supernovae yields of early stellar populations,  we find that all individual components of the subDLAs appears to be enriched by progenitor masses of $\lesssim30$~\Msol. The observed inhomogeneities between components can be reproduced by differences in the progenitor mass or supernova explosion energy.  As such, the observed chemical inhomogeneities between components can be explained by poorly mixed gas from different nucleosynthetic events.

\end{abstract}


\section{Introduction}

Our knowledge of the production and distribution of the first metals in the Universe is somewhat lacking, largely owing to the difficulty of finding and studying the first generation of stars in the Universe. Our current understanding mostly comes from observations of environments that are close to pristine, combined with detailed hydrodynamic simulations that aim to simulate the first stars from cosmological initial conditions (for a review on these topics, see \citealt{Glover2013,FrebelNorris2015,Klessen2019,Klessen23}).

From an observational standpoint, the chemical fingerprint of the first stars may be encoded in the abundance patterns of long-lived second generation stars that are in orbit around the Milky Way today. Dedicated surveys for such stars \citep{Bond1980,Beers1992,Christlieb2001,Cayrel2004,Bonifacio2009,Yong2013,Roederer2014,Starkenburg2017} have uncovered an incredible diversity of chemical abundance patterns at the metal-poor tail of the metallicity distribution function. Even though a metal-free star has not yet been identified from these works, the chemical fingerprints that have been measured are reproduced by models of metal-free stellar nucleosynthesis \citep{HegerWoosley2010,CookeMadau2014,Ishigaki2014,Marassi2014,Tominaga2014,Hartwig18,Welsh21,Hartwig23, Vanni23}. We have learnt a great deal about the first stars by modelling the chemistry of the putative second generation stars; while the details of the initial mass function shape are still uncertain, we have learnt that the dominant enrichment channel appears to be Type \textsc{ii} supernovae, some of which are faint and the minihaloes that host Population \textsc{iii} stars probably had a low multiplicity \citep{Clark08,Stacy10,Susa14,Jeon21}. Despite the successes of this approach, it is difficult to reliably measure the relative abundances of the most common chemical elements, and stellar modelling has several complexities that often result in model atmospheres that are computed in one-dimension and assume that lines form in regions of local thermodynamic equilibrium. Thus, complementary avenues to study the chemistry of the first stars are equally important.

The gaseous reservoirs in galactic environments, as seen in absorption along quasar (QSO) sightlines, are an important tool to study chemical evolution across a wide range of redshifts. In particular, damped Ly$\alpha$ systems (DLAs; \HI{} column densities log\,N(\HI{})$/{\rm cm}^{-2}\geq20.3$) and subDLAs ($19\leq$~log\,N(\HI{})$/{\rm cm}^{-2}<20.3$) contain the bulk of neutral gas across cosmic time \citep{Prochaska09,Noterdaeme12,Zafar13,SanchezRamirez16,Berg19} and are believed to be tracing interstellar or circumgalactic gas \citep{Wolfe05,Tumlinson17}. Given the high \HI\ column densities of subDLAs and DLAs, most of the elements reside in a dominant ionization state, allowing the chemical abundances to be measured robustly without the need for a significant accounting of unseen ion stages. Furthermore, the strong \HI\ absorption lines make it straightforward to efficiently identify gas clouds over a wide redshift range. For these reasons, DLAs and subDLAs allow us to study the metal enrichment of galaxies across cosmic time \citep{Rafelski12, Quiret16, Berg21}, and the detailed chemical evolution of galaxies. This approach to studying the chemical buildup of the Universe complements the efforts towards the same goals using the abundances of stars in the Milky Way \citep{Prochaska02,Berg15,Cooke15,Skuladottir18,Welsh19}. In particular, studies of very metal-poor (VMP) DLAs and absorbers provide insights into the chemical enrichment of the Universe from the first stars and galaxies \citep{Erni06,Pettini08,Penprase10,Cooke11,Welsh19,Saccardi23,Sodini24} and test theories on cold flow accretion to fuel star formation \citep{Fumagalli16a}.

There are pros and cons of using DLAs and subDLAs as tracers of detailed chemical enrichment over cosmic time. The main difficulty of using VMP DLAs to trace the evolution of first stars is that these absorbers are relatively rare on the sky, and the most abundant metals (namely C and O) are sometimes saturated in the VMP regime, particularly if the \HI\ column density is very high. On the other hand, systems with a relatively high \HI\ column density allows for a wider range of somewhat less abundant elements (e.g.~N, S, Ni) to be detected. SubDLAs have an incidence rate that is $\approx3\times$ higher than DLAs \citep{Zafar13,Berg19} and the C and O lines are not typically saturated. However, given the lower \HI\ column densities, the gas is less effective at self-shielding to ionizing radiation. As a result, subDLAs often require ionization corrections (ICs) to convert the measured ionic abundances to the intrinsic abundances \citep{Vladilo01, DessaugesZavadsky09,Milutinovic10,Fumagalli16b}. To date, the detailed chemistry of VMP subDLAs has only been studied at modestly high spectral resolution in a dozen absorbers \citep{DessaugesZavadsky03,Pettini08,Cooke11,Cooke14,Morrison16,Poudel20,Berg21,Saccardi23}. These studies have found the [C/O] and [Si/O] chemical abundances tend to agree with VMP DLA population, although [C/O] can be slightly higher in VMP subDLAs \citep{Poudel20,Berg21}. \cite{Morrison16} also noted that the individual absorption components of the metal-poor subDLA studied show variations in the abundance ratios of [C/O] and [Si/O]. These chemical variations between absorption components were predominantly attributed to differing ionizing conditions for each component, but \cite{Morrison16} speculate that the higher velocity component in their particular system could be explained by enrichment from a stellar outflow.

In this paper, we present UVES observations of five VMP subDLAs at redshift \zabs{}~$\approx4$ selected from the XQ-100 survey \citep{Lopez16,Berg21}, increasing the sample of metal-poor subDLAs observed at high spectral resolution by nearly 50\%. Using this sample, we investigate whether VMP subDLAs can be used to constrain the chemical abundance yields from the first generations of stars. In particular, we present a novel IC method to account for ionization effects in individual absorption components (Section \ref{sec:Ion}) in order to accurately assess abundance variations between these subDLA absorption components (Section \ref{sec:compAbund}). Lastly, we match the observed abundance ratios between components to the yields of core-collapse supernovae of metal-poor and pristine stars in order to search for differences in the properties of the progenitor population (Section \ref{sec:model}). Throughout the paper we assume the meteoritic \cite{Asplund09} solar abundance scale, with the exception of oxygen, where we assume the photospheric solar abundance scale from \cite{Asplund09}.

\section{Data}

\subsection{Sample selection, observations and data reduction}

Our sample of five VMP subDLAs were selected from the catalogue of subDLAs identified by the XQ-100 survey \citep{Berg21} for follow-up with the Ultraviolet and Visual Echelle Spectrograph \citep[UVES;][]{Dekker00} in order to resolve individual components and determine precise chemical abundances to draw meaningful comparisons with nucleosynthetic models. In short, the XQ-100 survey \citep{Lopez16} observed 100 redshift $3.5\leq$~\zem~$\leq4.5$ QSOs with X-Shooter (spectral resolving power $R\approx6000-9000$), providing an unbiased sample of subDLAs and DLAs. The VMP subDLA sample was chosen based on the following four criteria: (i) the measured metallicity\footnote{Throughout this paper, we use the square bracket notation, [X/Y], to denote the abundance of element X relative to Y, relative to the solar value: [X/Y]=log(X/Y)$-$log(X/Y)$_{\odot}$.} ([M/H]) of the subDLA from \cite{Berg21} was [M/H]~$<-2.5$, (ii) the C\ion{ii} $\lambda1334$\,\AA\ and O\ion{i} $\lambda1302$\,\AA\ lines are outside the Ly$\alpha$ forest or could easily be separated with blends with the Ly$\alpha$ forest, (iii) the XQ-100 data showed little or no contamination near the key metal ions of interest (i.e.~C\ion{ii}, O\ion{i}, Al\ion{ii}, Si\ion{ii}, and Fe\ion{ii}), and (iv) could be observed with a single observation setup with UVES.

\begin{table*}
\begin{center}
\caption{Summary of observations}
\label{tab:obs}
\begin{tabular}{lcccccc}
\hline
QSO& \zem\ & RA (J2000) & Dec. (J2000)& R (mag) & Exposure time (s) & S/N (pixel$^{-1}$)\\
\hline
J0034+1639& 4.290& 00:34:54.71& $+$16:39:18.2& 18.0& 14305& 22--31\\
J0415$-$4357& 4.070& 04:15:15.18& $-$43:57:50.7& 18.8& 65846& 15--23\\
J0529$-$3552& 4.170& 05:29:20.94& $-$35:52:31.8& 18.3& 14965& 18--25\\
J1032+0927& 3.990& 10:32:21.26& $+$09:27:47.5& 17.9& 14903& 21--28\\
J2251$-$1227& 4.160& 22:51:18.19& $-$12:27:05.1& 18.6& 14965& 9--15\\
\hline
\end{tabular}
\end{center}
\end{table*}

Our VMP subDLA sample was observed with the Very Large Telescope (VLT) UVES spectrograph  as part of programmes 94.A-0233(A) \citep{Berg16}, 104.A-0149(A), and 110.23QP (PI Berg). All quasars were observed with a minimum of 5 exposures of $\approx1$\,hr each with a $1''$ slit using the standard dichroic-2 setup in wavelength setting 437+760, with $2\times2$ binning; the wavelength range covered for each spectrum is: 3760-4990\AA{}, 5690-7525\AA{}, and 7650-9465\AA{}. For the QSO J0415$-$4357, we also included three exposures of $\approx1$\,hr using the red chip (wavelength setting 600) setup in order to cover the wavelength gap from the dichroic-2 setup with a signal-to-noise ratio S/N~$\approx8$ to confirm the \HI{} column density of the absorber using Ly$\alpha$ measured with X-Shooter data \citep{Berg19}. The nominal spectral resolution of this setup has a full-width at half maximum (FWHM) $\sim7.5~{\rm km~s}^{-1}$. The obtained individual spectra were reduced and extracted using the point source workflow of \textsc{ESO-reflex} \citep{ESOreflex}. The extracted individual spectra were then combined using \textsc{UVES\_popler} \citep{UVESpopler}. A first-order estimate of the QSO continuum was done by fitting a cubic spline to points of visually clean continuum within the combined spectrum. Table \ref{tab:obs} summarizes the properties of the quasar (QSO redshift $z_{\rm em}$, coordinates, and R-band magnitude), total exposure time for each sightline, and the interquartile range of S/N (per 1.8 [QSOs J0415$-$4357 and J2251$-$1227] or 2.5 km~s$^{-1}$ pixel [QSOs J0034+1639, J0529$-$3552, J1032+0927]) achieved for the combined spectrum beyond the Ly$\alpha$ emission of the QSO.

\subsection{Column density measurements}
\label{sec:cols}

Metal column density measurements for the detected absorption features were derived by Voigt profile fitting the data with the \textsc{alis} software \citep{Cooke14}. In summary, \textsc{alis} uses a chi-squared minimization to simultaneously fit the continuum and individual Voigt profile components. The fit is done on user-selected regions of the total absorption profiles. For each sightline, we simultaneously fit the absorption profiles of all the detected low ions (C\ion{ii}, \OI{}, Al\ion{ii}, Si\ion{ii}, and Fe\ion{ii}) separately from the high ions (C\ion{iv} and Si\ion{iv}). We typically used a third or fourth order Legendre polynomial for the continuum fit to account for any small variations missed by the first-order cubic spline continuum fit.  For each set of ions, we assumed a common velocity profile for the detected lines; specifically, each component shares the same turbulent Doppler parameter ($b$) and redshift ($z$) across all ions. We attempted to include thermal broadening in the fitting procedure, but we found that, given the quality of the data, we are unable to distinguish between thermally or turbulently broadened kinematics. To appropriately account for the effects of the different atomic mass of the ions, we assume a gas temperature of $10^4\,{\rm K}$ in line with other VMP subDLAs and DLAs \citep{Cooke15}. Changing this assumed temperature by $\pm2000\,{\rm K}$ has a minimal effect on the measured column densities; the typical difference in column density corresponds to a $\lesssim0.1\sigma$ difference, but can be as large as $\lesssim0.5\sigma$ for the weakest components. In cases where obvious blending can be identified within a single absorption line, additional components were included in the fit solely for that absorption line.

The individual Voigt profile component fitting parameters ($z$, $b$, and column density $N$) of the low and high ions are shown in Tables \ref{tab:loComp} and \ref{tab:hiComp} respectively, along with the relative velocity of the components to \zabs\ (\deltaV). Figures \ref{fig:J0034vel}--\ref{fig:J2251vel} show the \textsc{alis}-normalised data (black lines) and best-fitting model profiles (solid magenta curves) for the five VMP subDLAs of our sample; any additional blended absorption is represented by cyan curves. The significance of the column density detection is provided beside each column density in Tables \ref{tab:loComp} and \ref{tab:hiComp}. 

In order to include all the errors in the fitting procedure (i.e.~errors from continuum fitting, blending, and velocity profile parameters) whilst including upper limits, we opted to sample the covariance matrix outputted by \textsc{alis} in order to produce a representative sample of 10\,000 total column density measurements. The resampling of the component column densities is done in linear space (i.e. we use $N$, instead of log\,$N/{\rm cm}^{-2}$) and we ensure 10\,000 samples are drawn such that the component column densities have $N\geq0~{\rm cm}^{-2}$ to avoid non-physical parameters. The median total column densities of each ion (with errors based on the 25$^{\rm th}$/75$^{\rm th}$ percentiles of the 10\,000 samples) are provided in Table \ref{tab:Ntot} along with the background quasar's redshift ($z_{\rm em}$) and the \HI\ column density \citep[determined using X-Shooter data;][]{Berg19} of the absorber. The absorber redshift \zabs{} in Table \ref{tab:Ntot} is based on the strongest, central component of the low ions' (or C\ion{iv} for J0034+1639) absorption profile fits  described above. The total column densities measured from the UVES data presented here are generally consistent with those measured in the X-Shooter spectra \citep{Berg19}. There are some cases where the total column densities are inconsistent with the X-Shooter measurements by up to $\approx0.1$~dex, and can be attributed to differences in either: the continuum fitting (such as Si\ion{ii} 1304\AA{} towards J2251$-$1227), the absorption lines used (e.g.~Fe\ion{ii} 2600\AA{} [X-Shooter] compared to 1608\AA{} [UVES] towards J1032+0927), or identifying and removing blends in the higher-resolution UVES data (e.g.~C\ion{ii} 1334 in the J0415$-$4357 subDLA).

We summarize some of the specific details about the fitting procedure for each individual absorber below.

\begin{table}
\begin{center}
\caption{Component Voigt profile fits for low ions}
\label{tab:loComp}
\begin{adjustbox}{angle=90}
\begin{tabular}{lcccccccc}
\hline
Component& $z$& \deltaV{} (km s$^{-1}$)& $b$ (km s$^{-1}$)& logN(C\ion{ii})/${\rm cm}^{-2}$& logN(O\ion{i})/${\rm cm}^{-2}$& logN(Si\ion{ii})/${\rm cm}^{-2}$& logN(Al\ion{ii})/${\rm cm}^{-2}$& logN(Fe\ion{ii})/${\rm cm}^{-2}$\\
\hline
\multicolumn{9}{c}{\textbf{J0415$-$4357}}\\
1& $4.033553\pm0.000052$& $-92$& $15.9\pm4.71$& $12.76\pm0.166$ $[2.6\sigma]$& $12.65\pm0.294$ $[1.5\sigma]$& $11.64\pm0.168$ $[2.6\sigma]$& $11.24\pm0.186$ $[2.3\sigma]$& $12.70\pm0.211$ $[2.1\sigma]$\\
2& $4.034036\pm0.000006$& $-63$& $9.8\pm0.59$& $13.22\pm0.052$ $[8.4\sigma]$& $12.71\pm0.259$ $[1.7\sigma]$& $12.50\pm0.025$ $[17.6\sigma]$& $11.66\pm0.062$ $[7.0\sigma]$& $10.61\pm0.504$ $[0.9\sigma]$\\
3& $4.034572\pm0.000229$& $-32$& $43.0\pm16.09$& $12.98\pm0.197$ $[2.2\sigma]$& $12.80\pm0.403$ $[1.1\sigma]$& $11.80\pm0.202$ $[2.1\sigma]$& $10.57\pm0.546$ $[0.8\sigma]$& $12.34\pm0.889$ $[0.5\sigma]$\\
4& $4.035102\pm0.000003$& $0$& $5.1\pm0.32$& $12.76\pm0.351$ $[1.2\sigma]$& $13.15\pm0.178$ $[2.4\sigma]$& $12.46\pm0.051$ $[8.6\sigma]$& $11.23\pm0.148$ $[2.9\sigma]$& $12.22\pm1.281$ $[0.3\sigma]$\\
5& $4.036198\pm0.000013$& $65$& $15.0\pm1.14$& $12.75\pm0.105$ $[4.1\sigma]$& $13.46\pm0.046$ $[9.4\sigma]$& $12.21\pm0.031$ $[13.9\sigma]$& $10.98\pm0.310$ $[1.4\sigma]$& $12.77\pm0.139$ $[3.1\sigma]$\\
\hline
\multicolumn{9}{c}{\textbf{J0529$-$3552}}\\
1& $3.684225\pm0.000002$& $0$& $5.1\pm0.23$& \nodata{}& \nodata{}& $13.68\pm0.025$ $[17.5\sigma]$& $12.31\pm0.001$ $[324.2\sigma]$& \nodata{}\\
\hline
\multicolumn{9}{c}{\textbf{J1032+0927}}\\
1& $3.803621\pm0.000022$& $-14$& $5.7\pm2.22$& $12.90\pm0.099$ $[4.4\sigma]$& $13.01\pm0.205$ $[2.1\sigma]$& $11.69\pm0.508$ $[0.9\sigma]$& $11.07\pm0.134$ $[3.2\sigma]$& $11.82\pm0.553$ $[0.8\sigma]$\\
2& $3.803851\pm0.000002$& $0$& $4.7\pm0.21$& $14.00\pm0.027$ $[15.8\sigma]$& $14.65\pm0.032$ $[13.6\sigma]$& $13.31\pm0.022$ $[19.7\sigma]$& $11.80\pm0.029$ $[15.2\sigma]$& $13.02\pm0.048$ $[9.0\sigma]$\\
3& $3.804279\pm0.000040$& $27$& $18.2\pm2.85$& $12.84\pm0.085$ $[5.1\sigma]$& $13.19\pm0.515$ $[0.8\sigma]$& $12.18\pm0.315$ $[1.4\sigma]$& $11.11\pm0.182$ $[2.4\sigma]$& $12.15\pm0.434$ $[1.0\sigma]$\\
\hline
\multicolumn{9}{c}{\textbf{J2251$-$1227}}\\
1& $3.988500\pm0.000004$& $-7$& $0.5\pm0.00$& $13.18\pm0.088$ $[4.9\sigma]$& $12.46\pm0.590$ $[0.7\sigma]$& $12.25\pm0.078$ $[5.6\sigma]$& $10.88\pm0.388$ $[1.1\sigma]$& \nodata{}\\
2& $3.988616\pm0.000006$& $0$& $4.5\pm0.37$& $13.25\pm0.069$ $[6.3\sigma]$& $13.67\pm0.045$ $[9.6\sigma]$& $12.66\pm0.026$ $[16.4\sigma]$& $11.20\pm0.276$ $[1.6\sigma]$& \nodata{}\\
\hline

\end{tabular}
\end{adjustbox}
\end{center}
\end{table}

\begin{table*}
\begin{center}
\caption{Component Voigt profile fits for high ions}
\label{tab:hiComp}
\begin{tabular}{lccccc}
\hline
Component& $z$& \deltaV{} (km s$^{-1}$)& $b$ (km s$^{-1}$)& logN(C\ion{iv})/${\rm cm}^{-2}$& logN(Si\ion{iv})/${\rm cm}^{-2}$\\
\hline
\multicolumn{6}{c}{\textbf{J0034+1639}}\\
1& $4.222996\pm0.000013$& $-193$& $15.5\pm1.17$& $13.09\pm0.026$ $[16.5\sigma]$& $12.24\pm0.131$ $[3.3\sigma]$\\
2& $4.223542\pm0.000004$& $-161$& $10.7\pm0.30$& $13.53\pm0.010$ $[43.5\sigma]$& $12.90\pm0.029$ $[15.0\sigma]$\\
3& $4.226067\pm0.000033$& $-17$& $28.1\pm2.11$& $13.26\pm0.049$ $[8.9\sigma]$& $12.44\pm0.216$ $[2.0\sigma]$\\
4& $4.226052\pm0.000006$& $-17$& $4.8\pm0.82$& $13.05\pm0.046$ $[9.4\sigma]$& $12.50\pm0.084$ $[5.2\sigma]$\\
5& $4.226357\pm0.000004$& $0$& $6.7\pm0.47$& $13.39\pm0.024$ $[18.3\sigma]$& $12.89\pm0.048$ $[9.0\sigma]$\\
6& $4.227159\pm0.000005$& $46$& $5.4\pm0.64$& $13.03\pm0.038$ $[11.3\sigma]$& $12.41\pm0.067$ $[6.5\sigma]$\\
7& $4.227726\pm0.000017$& $79$& $8.9\pm2.49$& $12.84\pm0.205$ $[2.1\sigma]$& $12.37\pm0.156$ $[2.8\sigma]$\\
8& $4.227859\pm0.000028$& $86$& $30.6\pm2.95$& $13.86\pm0.018$ $[24.6\sigma]$& $12.59\pm0.141$ $[3.1\sigma]$\\
9& $4.228654\pm0.000023$& $132$& $14.9\pm0.97$& $13.58\pm0.043$ $[10.1\sigma]$& $12.13\pm0.362$ $[1.2\sigma]$\\
10& $4.228800\pm0.000004$& $140$& $6.9\pm0.37$& $13.69\pm0.036$ $[12.0\sigma]$& $13.25\pm0.026$ $[16.8\sigma]$\\
11& $4.229906\pm0.000017$& $204$& $13.8\pm1.68$& $12.71\pm0.048$ $[9.0\sigma]$& $11.24\pm1.150$ $[0.4\sigma]$\\
\hline
\multicolumn{6}{c}{\textbf{J0415$-$4357}}\\
1& $4.033482\pm0.000018$& $-96$& $12.6\pm1.24$& $12.82\pm0.063$ $[6.9\sigma]$& $12.70\pm0.057$ $[7.6\sigma]$\\
2& $4.034058\pm0.000013$& $-62$& $20.0\pm1.56$& $13.47\pm0.033$ $[13.3\sigma]$& $13.30\pm0.036$ $[12.1\sigma]$\\
3& $4.034389\pm0.000009$& $-42$& $5.2\pm1.75$& $12.49\pm0.163$ $[2.7\sigma]$& $12.42\pm0.135$ $[3.2\sigma]$\\
4& $4.035772\pm0.000028$& $40$& $57.1\pm3.75$& $13.64\pm0.030$ $[14.6\sigma]$& $12.97\pm0.051$ $[8.4\sigma]$\\
5& $4.034807\pm0.000012$& $-18$& $3.6\pm1.90$& $12.00\pm0.154$ $[2.8\sigma]$& $12.13\pm0.098$ $[4.5\sigma]$\\
6& $4.037134\pm0.000055$& $121$& $22.9\pm3.41$& $13.31\pm0.098$ $[4.4\sigma]$& $12.15\pm0.132$ $[3.3\sigma]$\\
7& $4.037454\pm0.000022$& $140$& $4.0\pm4.49$& $12.26\pm0.283$ $[1.5\sigma]$& $10.58\pm0.540$ $[0.8\sigma]$\\
8& $4.037712\pm0.000097$& $155$& $20.6\pm4.93$& $13.02\pm0.197$ $[2.2\sigma]$& $11.57\pm0.438$ $[1.0\sigma]$\\
9& $4.039217\pm0.000108$& $245$& $75.5\pm16.32$& $13.13\pm0.131$ $[3.3\sigma]$& $11.93\pm0.501$ $[0.9\sigma]$\\
10& $4.039607\pm0.000055$& $268$& $0.5\pm0.00$& $11.45\pm0.305$ $[1.4\sigma]$& $11.06\pm0.418$ $[1.0\sigma]$\\
\hline
\multicolumn{6}{c}{\textbf{J0529$-$3552}}\\
1& $3.682326\pm0.000009$& $-122$& $5.5\pm1.07$& \nodata{}& $12.30\pm0.043$ $[10.1\sigma]$\\
2& $3.682835\pm0.000007$& $-89$& $6.5\pm0.75$& \nodata{}& $12.52\pm0.029$ $[14.8\sigma]$\\
3& $3.683529\pm0.000012$& $-45$& $3.9\pm2.24$& \nodata{}& $12.28\pm0.149$ $[2.9\sigma]$\\
4& $3.683673\pm0.000021$& $-35$& $15.5\pm1.11$& \nodata{}& $12.84\pm0.049$ $[8.9\sigma]$\\
5& $3.684175\pm0.000012$& $-3$& $0.5\pm0.00$& \nodata{}& $11.94\pm0.075$ $[5.8\sigma]$\\
6& $3.685332\pm0.000017$& $71$& $11.7\pm1.70$& \nodata{}& $12.37\pm0.050$ $[8.7\sigma]$\\
\hline
\multicolumn{6}{c}{\textbf{J1032+0927}}\\
1& $3.803621\pm0.000012$& $-14$& $11.9\pm1.14$& $13.09\pm0.035$ $[12.5\sigma]$& $12.12\pm0.075$ $[5.8\sigma]$\\
2& $3.804279\pm0.000040$& $27$& $18.2\pm2.85$& $12.97\pm0.086$ $[5.1\sigma]$& $12.45\pm0.090$ $[4.9\sigma]$\\
3& $3.804532\pm0.000023$& $42$& $7.6\pm2.84$& $12.46\pm0.226$ $[1.9\sigma]$& $12.06\pm0.189$ $[2.3\sigma]$\\
4& $3.804959\pm0.000020$& $69$& $11.1\pm2.09$& $12.26\pm0.134$ $[3.2\sigma]$& $12.31\pm0.066$ $[6.6\sigma]$\\
\hline
\multicolumn{6}{c}{\textbf{J2251$-$1227}}\\
1& $3.988028\pm0.000017$& $-35$& $1.2\pm6.29$& $12.34\pm0.117$ $[3.7\sigma]$& $11.92\pm0.163$ $[2.7\sigma]$\\
2& $3.988423\pm0.000021$& $-12$& $12.1\pm2.12$& $12.79\pm0.059$ $[7.4\sigma]$& $12.37\pm0.095$ $[4.6\sigma]$\\
\hline

\end{tabular}
\end{center}
\end{table*}

\begin{table}
\begin{center}
\caption{Uncorrected total column densities of the VMP subDLA sample}
\label{tab:Ntot}
\begin{adjustbox}{angle=90}
\begin{tabular}{lccccccccccc}
\hline
QSO& $z_{\rm em}$& $z_{\rm abs}$& logN(\HI)& logN(C\ion{ii})& logN(O\ion{i})& logN(Al\ion{ii})& logN(AlI\ion{iii})& logN(Si\ion{ii})& logN(Fe\ion{ii})& logN(C\ion{iv})& logN(Si\ion{iv})\\
\hline
J0034+1639& 4.290& 4.22636& $19.60\pm0.20$& $<12.50$& \nodata{}& $<11.13$& \nodata{}& $<12.29$& $<12.85$& $14.45_{-0.003}^{+0.003}$& $13.71_{-0.011}^{+0.011}$\\
J0415$-$4357& 4.070& 4.03510& $20.05\pm0.30$& $13.64_{-0.024}^{+0.022}$& $13.77_{-0.029}^{+0.026}$& $11.97_{-0.043}^{+0.037}$& \nodata{}& $12.94_{-0.010}^{+0.010}$& $13.17_{-0.070}^{+0.059}$& $14.12_{-0.014}^{+0.013}$& $13.61_{-0.014}^{+0.012}$\\
J0529$-$3552& 4.170& 3.68422& $20.10\pm0.20$& \nodata{}& \nodata{}& $12.31_{-0.001}^{+0.001}$& $<11.77$& $13.68_{-0.017}^{+0.017}$& \nodata{}& \nodata{}& $13.24_{-0.011}^{+0.011}$\\
J1032+0927& 3.990& 3.80385& $20.00\pm0.15$& $14.06_{-0.015}^{+0.015}$& $14.67_{-0.021}^{+0.020}$& $11.94_{-0.024}^{+0.022}$& $<12.69$& $13.35_{-0.019}^{+0.017}$& $13.10_{-0.035}^{+0.030}$& $13.42_{-0.016}^{+0.016}$& $12.86_{-0.022}^{+0.021}$\\
J2251$-$1227& 4.160& 3.98862& $19.60\pm0.15$& $13.52_{-0.026}^{+0.024}$& $13.70_{-0.028}^{+0.027}$& $11.37_{-0.117}^{+0.089}$& $<12.74$& $12.80_{-0.004}^{+0.004}$& \nodata{}& $12.92_{-0.033}^{+0.030}$& $12.50_{-0.055}^{+0.049}$\\
\hline
\end{tabular}
\end{adjustbox}
\end{center}
\end{table}

\subsubsection*{J0034+1639}

\begin{figure*}
\begin{center}
\includegraphics[width=\textwidth]{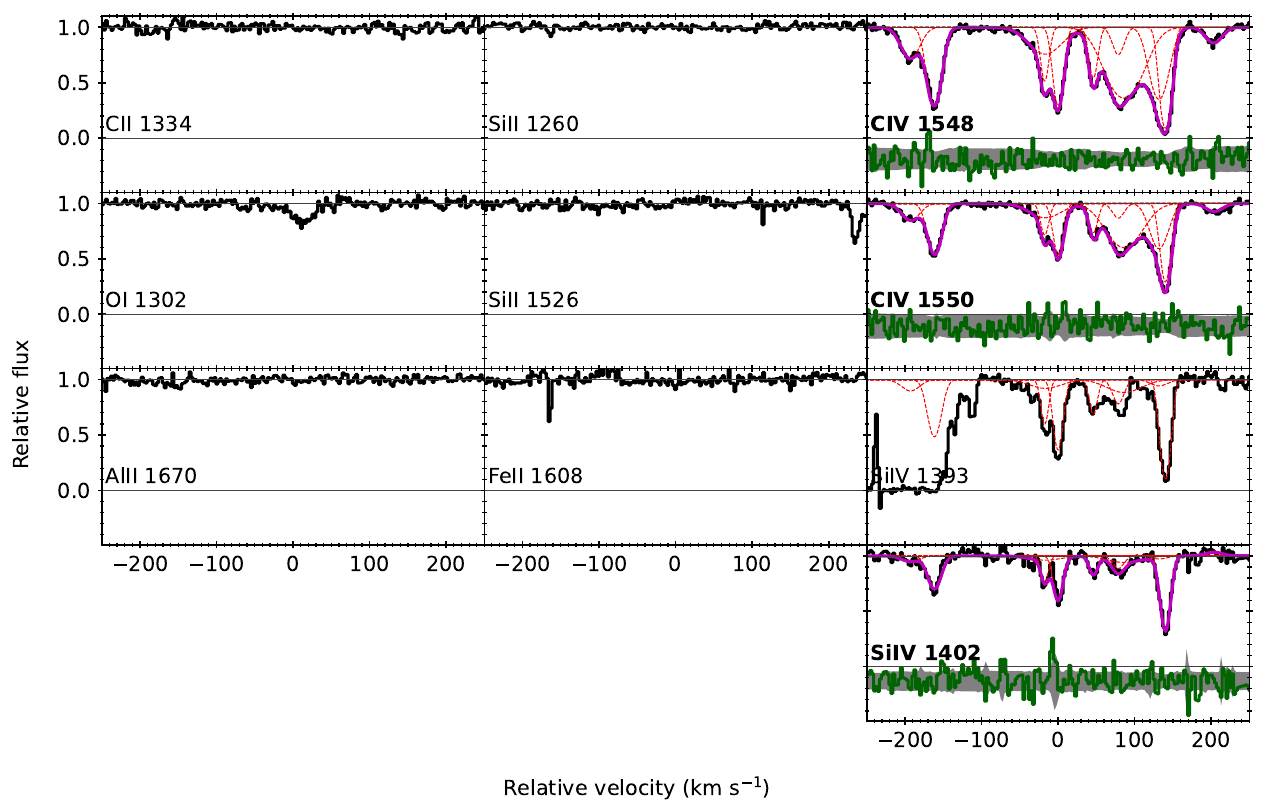}
\caption{Velocity profiles of the absorber towards QSO J0034+1639 (\zabs{}$=4.22636$). The black histogram shows the continuum-normalized UVES data. Panels showing absorption lines with bolded labels indicate the data that were used in the Voigt profile fitting analysis, with cyan curves to denote assumed blends included as part of the fit.  The solid magenta curve shows the overall Voigt profile fit if absorption was detected for that ion. The dashed red curves show the Voigt profile models of each component of the fit that was adopted as part of the total column density. The thin horizontal lines indicates the zero and continuum level. The residual from the fitting process is denoted by the green histogram at the bottom of the panel, with the shaded grey region indicating the $\pm1\sigma$ range of the residual based off the error spectrum. Both the residual curve and $\pm1\sigma$ region have been rescaled for display purposes. }
\label{fig:J0034vel}
\end{center}
\end{figure*}

\begin{figure*}
\begin{center}
\includegraphics[width=\textwidth]{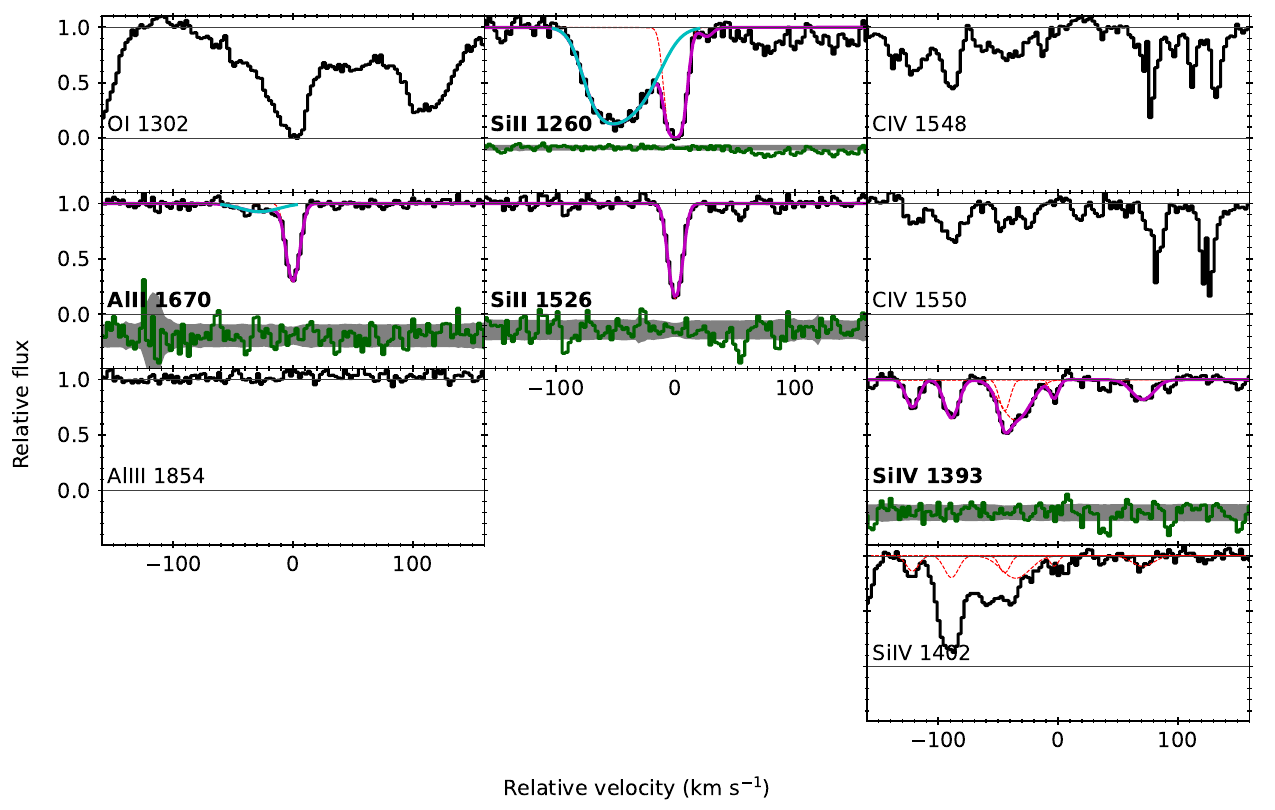}
\caption{Velocity profiles of the absorber towards QSO J0529$-$3552 (\zabs{}~$=3.68422$), following the same notation as in Figure \ref{fig:J0034vel}. }
\label{fig:J0529vel}
\end{center}
\end{figure*}

\begin{figure*}
\begin{center}
\includegraphics[width=\textwidth]{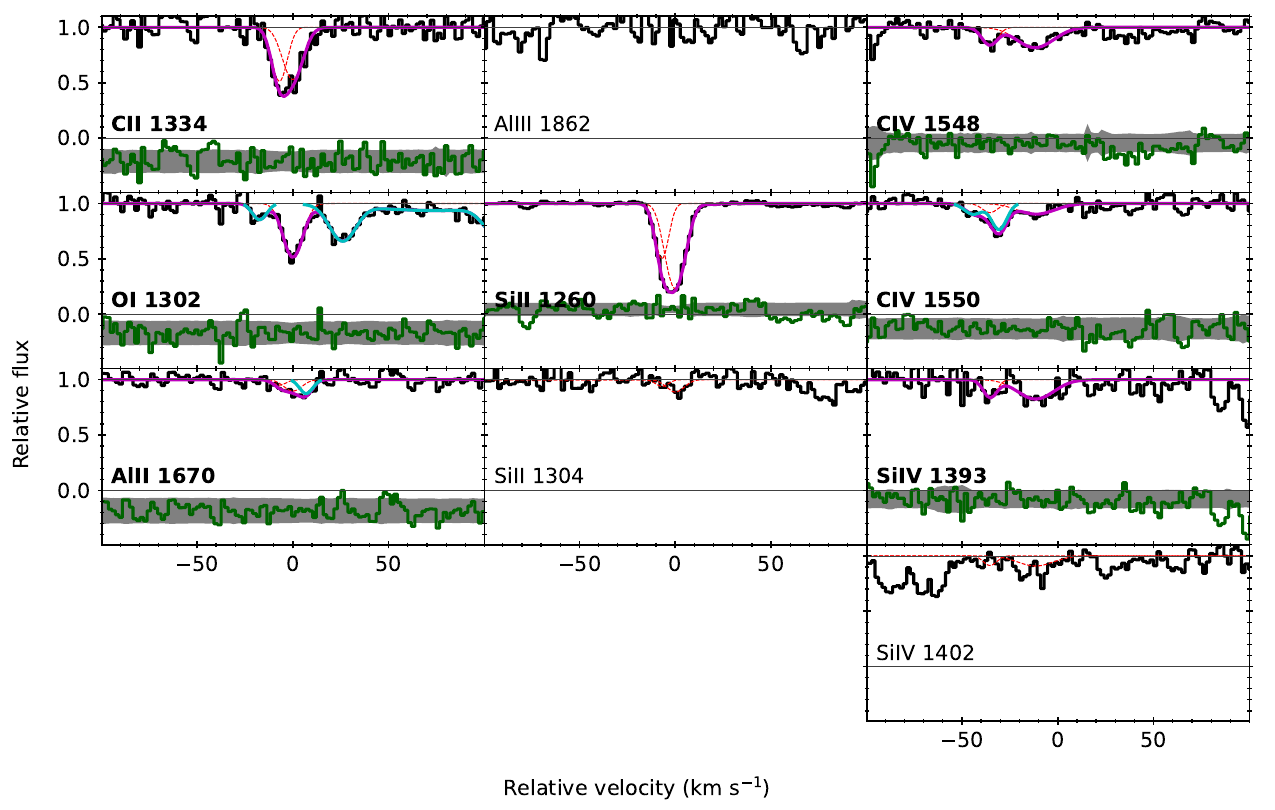}
\caption{Velocity profiles of the absorber towards QSO J2251$-$1227 (\zabs{}~$=3.98862$), following the same notation as in Figure \ref{fig:J0034vel}. }
\label{fig:J2251vel}
\end{center}
\end{figure*}

While there is significant \HI{}, C\ion{iv}, and Si\ion{iv} detected at redshift \zabs{}~$=4.227$, no low ionization metal lines appear to be detected, apart from a broad feature near the expected \OI{} 1302\AA{} line (Figure \ref{fig:J0034vel}). Assuming the absorption is indeed \OI{} 1302\AA{}, fitting a single Voigt profile implies that log~$N$(\OI{})/cm$^{-2}=13.55\pm0.05$, $b=17.4\pm2.0$, and $z=4.226567\pm0.000022$. The corresponding estimate of the metallicity is [O/H]~$=-2.7\pm0.2$. We note that using $b=17.4$~\kms{} and the S/N at the C\ion{ii} 1334\AA{} line, the estimated $3\sigma$ upper limit on [C/H] and [C/O] are [C/H]~$<-2.96$ and [C/O]~$<-0.26$ \citep[using the method to compute upper limits outlined in][]{Pettini94}. Based on \textsc{Cloudy} modelling (see Section \ref{sec:Ion}), the expected IC for [C/O] is $\approx-0.2$ at log~N(\HI)/cm$^{-2}=19.6$. Factoring in this IC results in the limit [C/O]~$\lesssim-0.46$, which is inconsistent with the lowest [C/O] detected in VMP subDLAs and VMP DLAs \citep{Cooke15,Berg21}.  We checked to see if the feature at \OI{} 1302\AA{} corresponds to a lower redshift absorber by assuming it was part of a strong doublet (Mg\ion{ii}, Ca\ion{ii}, Al\ion{iii}, Na\ion{i}) or a strong Fe\ion{ii} line (2344, 2382, or 2600\AA{}). While most of these lines are ruled out based on the lack of detection of another line, we cannot rule out the possibility that it is Mg\ion{ii} 2796\AA{}, Ca\ion{ii} 3934\AA{}, or Na\ion{i} 5897\AA{} from a lower redshift absorber, as their corresponding doublet feature would be blended with lines from other strong absorbers along the sightline \citep{Berg19}. We conservatively exclude this \OI{} line from our analysis as we cannot confirm if this absorption is indeed \OI{} at the redshift of the subDLA, and the suggested [C/O] upper limit is likely inconsistent with the rest of the typical VMP absorber population.  However, this subDLA is part of a complex system of proximate absorbers; the subDLA is $\approx5000$~\kms{} from the Ly$\alpha$ emission of the QSO and $\lesssim 3000$~\kms{} from two proximate DLAs. We caution that the combination of multiple absorbers \citep[i.e.~effects from star formation histories within group environments][]{Lopez03} or proximity effects \citep{Vladilo01, Rix07} can potentially lead to anomalous abundance patterns in strong Ly$\alpha$ absorbers \citep{Ellison10, Berg16}.  The $3\sigma$ upper limits on the total column densities adopted for the low ions for this absorber (Table \ref{tab:Ntot}) are derived using the \cite{Pettini94} method based on the maximum column density of an absorption line (with FWHM equal to the UVES resolution element) marginally detectable given the S/N within the velocity range shown in Figure \ref{fig:J0034vel}.

\subsubsection*{J0415$-$4357}
\begin{figure*}
\begin{center}
\includegraphics[width=\textwidth]{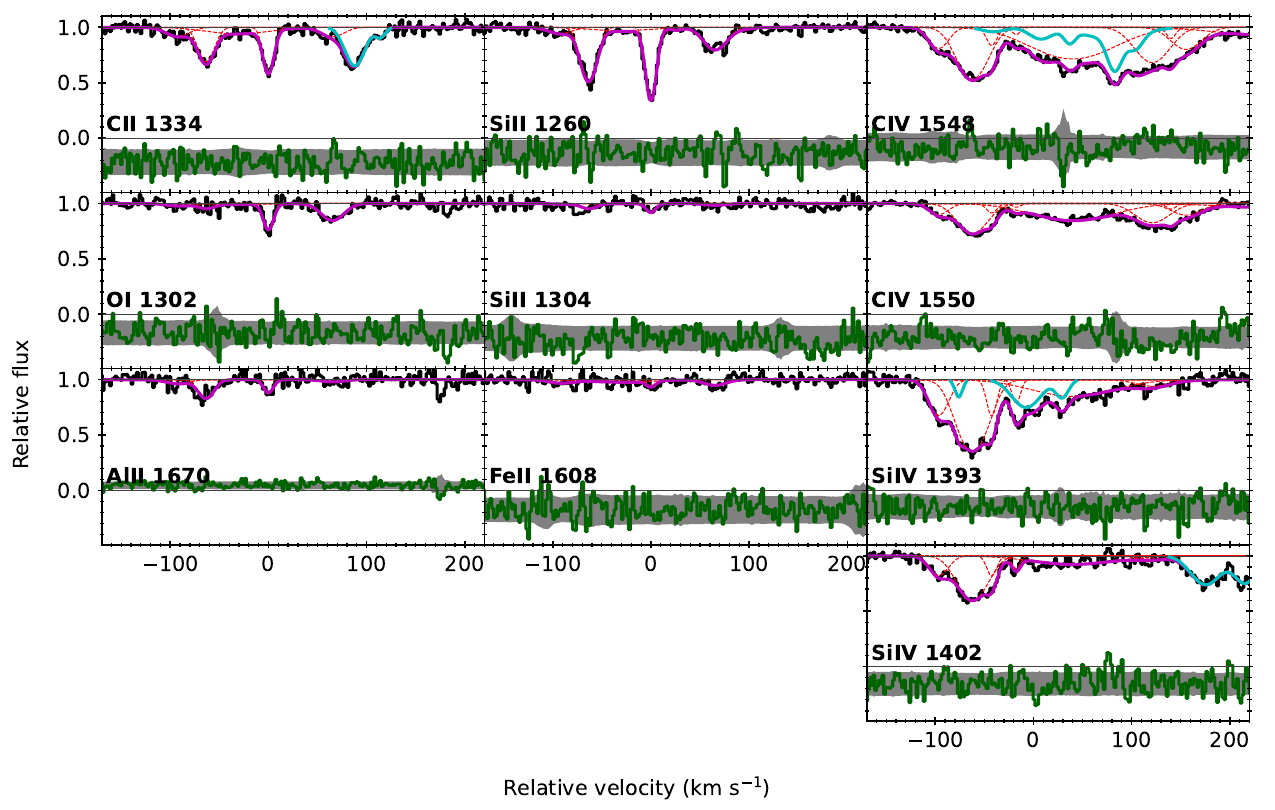}
\caption{Velocity profiles of the absorber towards QSO J0415$-$4357 (\zabs{}~$4.03510$), following the same notation as in Figure \ref{fig:J0034vel}. }
\label{fig:J0415vel}
\end{center}
\end{figure*}

The absorption profile shape of the low ionization species of the \zabs{}~$=4.03473$ absorber towards J0415$-$4357 is intriguing as there are three well-separated sets of components (Figure \ref{fig:J0415vel}). Each of these three sets of components show varying relative line strengths between different ions, such as nearly identical strength C\ion{ii} 1334\AA{} in the bluemost and central sets of components (i.e.~$\approx-70$ and $\approx0$~\kms{}, respectively) whilst the bluemost component set in O\ion{i} 1302 \AA{} is much weaker than the central component. We note that the fitting procedure prefers a very broad but weak component at a relative velocity of $-32$~\kms{}  over including additional curvature in the continuum fit in order to explain the small dip in relative flux in the two strongest lines C\ion{ii} 1334\AA{} and Si\ion{ii} 1260\AA{}. The redmost component of the C\ion{ii} 1334\AA{} line is blended with an intervening C\ion{iv} doublet at \zabs{}~$=3.34$. This blend was modelled using the C\ion{iv} 1550\AA{} line at \zabs{}~$=3.34$. The same method was applied to modelling the doublet blends in the C\ion{iv} 1548\AA{} profile, which is also blended with intervening systems at \zabs{}~$=4.027$.

\subsubsection*{J0529$-$3552}
While \OI{} 1302 is visually detected in the absorber at redshift \zabs{}~$=3.6836$ towards J0529$-$3552, this line is significantly blended with the Ly$\alpha$ forest. In our attempts to fit this \OI{} line, we tried to include fits to the Ly$\alpha$ by first estimating the column density and Doppler parameter from Ly$\beta$ absorption in the available XQ-100 data (as the Ly$\beta$ falls in the UVES chip gap of the dichroic set-up) and including these estimates in the fitting procedure. Unfortunately, the modeling is too degenerate given the uncertain continuum and Ly$\alpha$ absorption, leading to a poor estimate of the \OI{} column density. We therefore do not include an \OI{} column density measurement of this absorber.

\subsubsection*{J1032+0927}
\begin{figure*}
\begin{center}
\includegraphics[width=\textwidth]{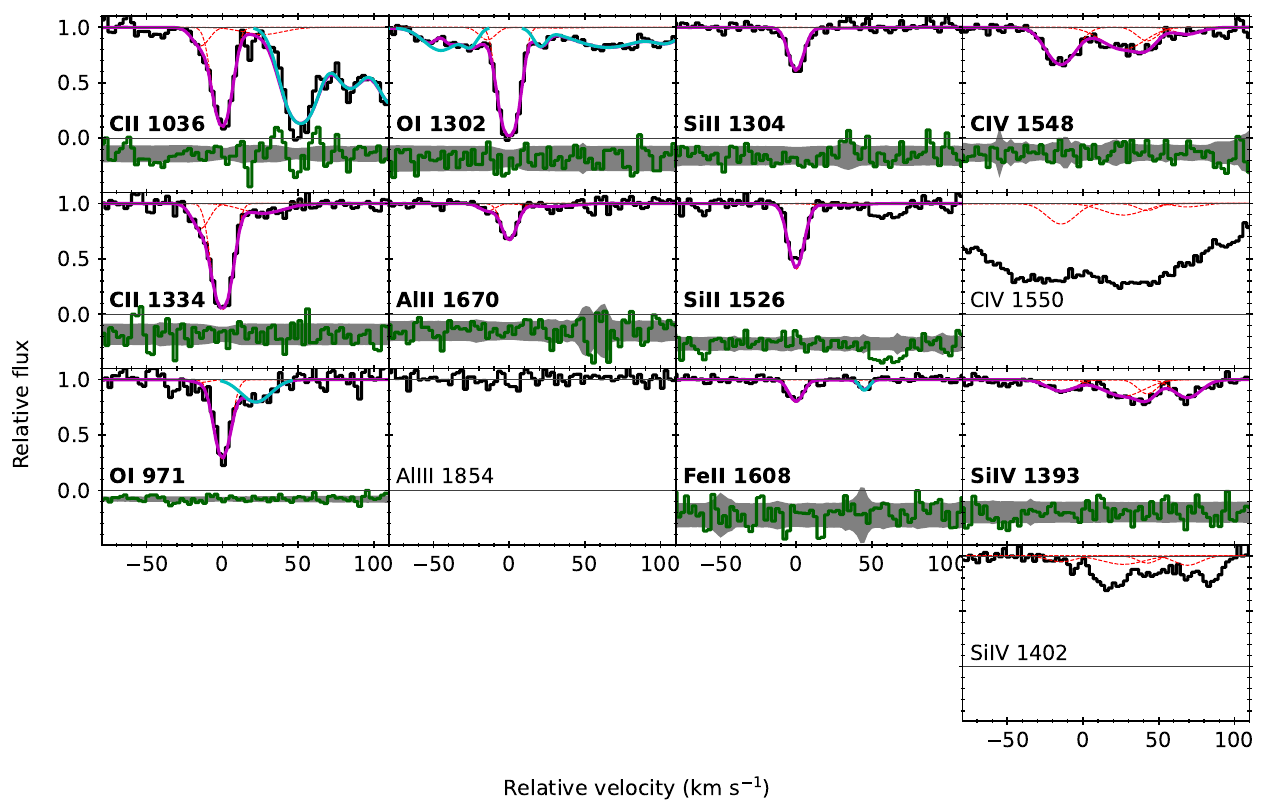}
\caption{Velocity profiles of the absorber towards QSO J1032+0927 (\zabs{}~$=3.80385$), following the same notation as in Figure \ref{fig:J0034vel}. We point out that the \OI{} 971\AA{} line sits within a broad feature of the Ly$\alpha$ forest. The Ly$\alpha$ forest component has been included as part of the continuum fit, and thus the forest line does not appear in the corresponding panel. }
\label{fig:J1032vel}
\end{center}
\end{figure*}

The O\ion{i} 971 \AA{} line is significantly blended with a Ly$\alpha$ forest line. As the bulk of this \OI\ absorption lies within a relatively clean section of the wing of the Ly$\alpha$ absorber, we treated the Ly$\alpha$ absorption as part of the continuum in the fit, and as a result is not seen in the O\ion{i} 971 \AA{} panel in Figure \ref{fig:J1032vel}. Based on the strongest metal absorption lines (i.e.~C\ion{ii} 1036\,\AA\ and 1334\,\AA), there appears to be at most three components in the low ions' velocity profiles. However, the components at \deltaV{}~$=-14$~\kms{} and $+27$~\kms{} are not detected in the Si\ion{ii} lines, and are only marginally detected in Al\ion{ii}. Unfortunately, the \OI{} lines are sufficiently blended that we are unable to improve upon the precision of the fits of these two components. We are confident the component at \deltaV{}~$=-14$~\kms{} is real based on the C\ion{ii} and Al\ion{ii} lines. The lack of asymmetry in the Si\ion{ii} 1304\AA{} and 1526\AA{} lines (and potentially O\ion{i} 971\AA{} and 1302\AA{}) suggests that ionization effects or nucleosynthetic differences between components may be important to consider. However, we caution that the component at \deltaV{}~$+27$~\kms{} requires additional confirmation with higher S/N data. Hereafter, we use all three components in our analysis.

\subsubsection*{J2251$-$1227}
Of the three strongly-detected metal lines (C\ion{ii} 1334 \AA{}, O\ion{i} 1302\AA{}, and Si\ion{ii} 1260 \AA{}) for the subDLA towards J2251$-$1227, only the C\ion{ii} absorption shows a significantly asymmetric profile that requires two components for fitting. While there is a hint of an asymmetry in the Si\ion{ii} 1260\AA{} line, it is not as strong as the C\ion{ii} asymmetry and can still be well fit by a single component. Unfortunately, the only other strong absorption line (O\ion{i} 1302\AA{}) is significantly blended, and cannot be used to conclude if one or two components are needed.  We investigated whether the C\ion{ii} line is blended with metal lines from an absorber at a different redshift as was done for J0415$-$4357, but we could not identify any additional absorption. While we caution the C\ion{ii} 1334\AA{} line is potentially blended, we continue to use a two component fit to this absorber.

\subsection{Accounting for ionization effects}
\label{sec:Ion}

While ICs are expected to be small for subDLAs, especially for the three systems in this work with log\,$N$(\HI{})/cm$^{-2}\geq20.0$ \citep{Vladilo01}, two of the absorbers (J0034+1639 and J0415$-$4357) are proximate to their host quasar ($\Delta V<5000$~\kms{}), and thus may be irradiated by a harder radiation field than the typical ultraviolet background of quasars and galaxies at this redshift. Furthermore, based on the information provided in Table \ref{tab:loComp}, we find that the ratios of any two ions attributed to a single component in a given system shows a variation; the ion column density ratios of the various components do not appear to be constant. While this could be a nucleosynthetic signature, it is possible that part or all of the signature is due to ionization effects \citep[e.g.][]{Morrison16}. It is therefore challenging to determine accurate ICs for the VMP subDLAs in our sample.

In this paper, we define the IC as the deficit between the intrinsic (${\rm log\,N_{int}}$) and observed (${\rm log\,N_{obs}}$) column density,
\begin{equation}
{\rm IC = log\,N_{int} - log\,N_{obs}.}
\label{eq:IC}
\end{equation}

Determining an accurate IC hinges on having an accurate model of the geometry of the gas as well as the co-location of ions within the model. Particularly for the case of J0415$-$4357 where several components of the low ions are well-separated in velocity space, it is unclear what sort of geometry best describes this system (e.g. the component clouds could be well-separated in physical space, or some components may be partly shielding other clouds from the nearby QSO or background radiation).  

\subsubsection*{Ionization modelling method}

To do such a detailed comparison of ionization-corrected abundance ratios in individual components using standard \cloudy{} \citep[][version C17.03]{Cloudy17} ionization modelling techniques, we would require the individual \HI{} column densities attributed to each component. In principle, one could estimate the \HI{} column density of each component by either assuming a constant metallicity (particularly [O/H], where the ionization potential of O\ion{i} is most similar to \HI{}) or measuring the column density of several ions and include the \HI{} column density of the component as an input model parameter (constrained by the total \HI{} column density of the system) and fitting all components simultaneously. Given the few confident (i.e.~$>5\sigma$) O\ion{i} column density measurements in these systems, additional constraints would be needed to accurately assess the typical ionization model parameters in addition to the \HI{} column density. However, the assumptions in both methods could wash out any nucleosynthetic differences between components.

Instead, we investigate a novel approach to use \cloudy{} ionization modelling on a component-by-component basis using an assumption that is independent of nucleosynthetic differences between components and estimate the ICs per component. Rather than assuming a constant metallicity for each component of the subDLA, we elected to start with the assumption that the intrinsic [Si/O] value of each component in VMP subDLAs is identical to the value seen in VMP DLAs \citep[${\rm [Si/O]_{DLA}=-0.15\pm0.15}$;][]{Cooke15,Welsh19} to derive the ICs for each component. [Si/O] is chosen because: (1) there is little scatter seen in the VMP DLAs; (2) both Si and O are $\alpha$ elements and have similar nucleosynthetic origins; and (3) Si\ion{ii} is sensitive to ionization effects, while \OI{} is expected to accurately trace the \HI\ gas. The idea behind this assumption is essentially a proof by contradiction. We start by assuming that there are no nucleosynthetic differences in [Si/O] across all components of a given absorber. If we start with this assumption, calculate and apply the ICs to the other ions, and find significantly different component abundances, then our initial assumption that the components have the same intrinsic abundance (i.e. the assumption that [Si/O] is the same for all components) is incorrect. On the other hand, if we find that all component abundances are mutually consistent with a constant value, then we cannot distinguish whether or not there is intrinsic variation among the components. The best we can do in this latter case is appeal to Occam's razor and conclude that the abundances are most likely drawn from a constant value, to within the errors of the data. In this section we summarize the method to derive the ICs and a simple validation of this method using a system with known \HI{} column densities in each component. In Section \ref{sec:compAbund} we revisit and discuss the implications of this assumption in the context of the results.

Based on this assumption of a constant [Si/O] within all clouds of a subDLA, we derive individual component ICs using a Markov chain Monte Carlo (MCMC) analysis of a grid of \cloudy{} models.  Using the measured log\,$N$(Si\ion{ii}) and log\,$N$(O\ion{i}) of each component as input (Table \ref{tab:loComp}) and ${\rm [Si/O]_{DLA}}$ as a constraint, we use \textsc{emcee} \citep{emcee} to find the distribution of required ICs for each component that are necessary to reproduce the observed [Si/O] in VMP DLAs.

We start by generating a grid of \cloudy{} models, assuming a plane-parallel geometry illuminated by a \cite{Haardt12} ultraviolet background at the redshift of the absorber, and the (uniformly scaled) \cite{Asplund09} solar abundances. We note that choosing either the \cite{Asplund09} scale or an abundance pattern similar to metal-poor stars, such as the \cite{Cayrel2004} abundance pattern as implemented by \cite{Saccardi23}, does not significantly change our derived ICs. As some of our systems are proximate to the host QSO (redshifts within $<5000$ \kms{}), we also include an AGN component to the radiation model in case there is additional contribution to the ionization field from the QSO. We use the AGN \cloudy{} function with the suggested parameters from the \cloudy{} documentation to represent a typical AGN continuum\footnote{The \cloudy{} AGN parameters adopted are: ``Big Bump'' temperature of $1.5\times10^5$K, X-ray to UV ratio of $-1.4$, ``Big Bump'' slope of $-0.5$, and X-ray slope of $-1.0$.}, and vary the ionization parameter ($U$). The final \cloudy{} grid varies the metallicity (\Zcloudy{}), volume density of hydrogen (\nH{}), N(\HI{}),  $z$, and $U$. Table \ref{tab:CloudyGrid} summarizes the range and step size of the parameters of the \cloudy{} grid. The step size was chosen to reduce the size of the grid while minimizing interpolation errors in the column densities for a proposed set of parameters.

\begin{table}
\caption{Range of \cloudy{} grid parameters}
\label{tab:CloudyGrid}
\begin{tabular}{lccc}
\hline
Parameter & Min. Value& Max. Value & Step size\\
\hline
$z$ & 3.75 & 4.75 & 0.25 \\
log\,N(\HI{})/${\rm cm}^{-2}$ & 17.0 & 20.5 & 0.25 \\
log(\nH{}) & $-4.0$ & 0.0 & 0.25 \\
\Zcloudy{} & $-3.9$ & $-1.5$ & 0.3 \\
log($U$) & $-8.0$ & 0.0 & 0.5 \\
\hline
\end{tabular}

\end{table}

The MCMC sampler proposes a set of \cloudy{} model parameters \nH{}, \Zcloudy{}, log\,N(\HI{})/${\rm cm}^{-2}$ for each component simultaneously. Additionally, a single log($U$) is proposed by the MCMC sampler and adopted for all components. For the non-proximate system, we force log($U$)~$=-8.0$ as the AGN ionizing radiation contribution is negligible below log($U$)~$\leq-5.0$. We assume a flat prior on log\,N(\HI{})/${\rm cm}^{-2}$ of each component to be between 17.0 and the total log\,N(\HI{})/${\rm cm}^{-2}$ of the system (Table \ref{tab:Ntot}).  For the proximate absorber J0415$-$4357, we assume a flat prior of $-5.0\leq$~log($U$)~$\leq0.0$ to ensure there is at least a minimal contribution from the QSO to the ionizing radiation. For the remaining parameters, we assume a flat prior within the range defined by the \cloudy{} model grid (Table \ref{tab:CloudyGrid}). We further require that the MCMC sampler only accepts proposed \cloudy{} models where the proposed \HI{} column densities of all components sum to the total \HI{} of the system (within the errors quoted in Table \ref{tab:Ntot}). We note that the errors on log\,N(\HI{})/${\rm cm}^{-2}$ of the subDLAs (and thus the dominant source of errors in the metallicities) are the upper and lower bounds on the total \HI{} column.

If the proposed model parameters are accepted, the \cloudy{} grid is re-interpolated to the proposed parameters at the redshift of the absorber. Using the proposed \cloudy{} parameters, the ICs (\IC{}) for both Si\ion{ii} and O\ion{i} in all components are derived using

\begin{equation}
\rm IC(X_{i}) = [M/H]_{cld} + logN(\HI)_{comp} + log(X/H)_{\odot} - logN_{cld}(X_{i}),
\label{eq:ICcloudy}
\end{equation}

where ${\rm log(X/H)_{\odot}}$  and ${\rm logN_{cld}(X_{i})}$ are the respective solar value and the \cloudy{} model's column density for ion X$_{\rm i}$. Using randomly-drawn values of the observed logN(Si\ion{ii}) and logN(O\ion{i}) for each component from the \textsc{alis} covariance matrix, the IC-corrected [Si/O] (${\rm[Si/O]_{IC}}$) is computed for each component using the derived \IC{} with Equation \ref{eq:IC}. The adopted likelihood function for each component in the MCMC analysis is:

\begin{equation}
\rm log(\mathcal{L}_{comp}) = \frac{-([Si/O]_{IC} - [Si/O]_{DLA})^2}{2\sigma_{DLA}^{2}}  - 0.5log(2 \pi \sigma_{DLA}^{2})
\end{equation}

where ${\rm \sigma_{DLA}}$ is the error on ${\rm [Si/O]_{DLA}}$ (i.e.~${\rm [Si/O]_{DLA}=-0.15\pm0.15}$). The total likelihood function for the MCMC sampler is the sum of $\rm log(\mathcal{L}_{comp})$ for all components. For each subDLA, we run the MCMC sampler for 3600 steps with 400 walkers. We adopt a conservative burn-in of 3300 steps. Although the bulk of the walkers appear to converge by $\lesssim1200$ steps in all systems, there are a non-negligible number of walkers that take the full 3300 steps to converge. Since the observed logN(Si\ion{ii}) and logN(O\ion{i}) are randomly drawn at each evaluation of the likelihood function from the covariance matrix (see Section~\ref{sec:cols}), the statistical errors of the measured column densities will be properly propagated within the MCMC likelihood computation given our large number of samples.  For each of the resulting 120\,000 proposed sets of \cloudy{} model parameters after burn-in, we measure the \IC{} for all low-ionization species in each component using Equation \ref{eq:ICcloudy}. Table \ref{tab:ICcomp} provides the median \cloudy{} model parameters and \IC{} values of each component; the adopted errors are calculated as the difference of the median with the 25$^{\rm th}$/75$^{\rm th}$ percentiles.

To obtain an IC on the total column density (i.e.~the `Combined' rows of Table \ref{tab:ICcomp}), we take each of the 10\,000 covariance-sampled N(Si\ion{ii}) and N(O\ion{i}) of each component and find the closest matching \cloudy{} model parameter set from the 120\,000 sets that reproduces these column densities by finding the minimum of ${\rm \Delta N^{2} = \Delta N(Si\ion{ii})^2 + \Delta N(O\ion{ii})^2}$. The matched models' ${\rm \Delta N}$ is typically $<0.05$~dex.  If there are multiple close matches, a single set is randomly chosen assuming a uniform distribution. Using these 10\,000 model sets, the resulting total IC per absorber is derived using Equation \ref{eq:IC}; logN$_{\rm int}$ is obtained by summing the model set's ionization-corrected column density for each component in linear space, and logN$_{\rm obs}$ is the corresponding total column density from summing the components of the corresponding covariance sample. Comparing the ICs for the individual components to the `Combined' value in Table \ref{tab:ICcomp} suggests that there can be significant variations in the ionization state across absorption components in our VMP subDLA sample.

\begin{table*}
\begin{center}
\caption{\cloudy{} model parameters and ICs for each low ion component}
\label{tab:ICcomp}
\begin{tabular}{lccccccccc}
\hline
Component& logN(\HI{})& log(\nH{})& \Zcloudy{}& log(U)& IC(C\ion{ii})& IC(O\ion{i})& IC(Si\ion{ii})& IC(Al\ion{ii})& IC(Fe\ion{ii})\\
\hline
\multicolumn{10}{c}{\textbf{J0415$-$4357}}\\
1& $19.3^{+0.02}_{-0.34}$& $-0.7^{+0.11}_{-0.00}$& $-3.5^{+0.19}_{-0.00}$& $-3.4^{+0.01}_{-0.01}$& $-0.43^{+0.006}_{-0.269}$& $-0.13^{+0.078}_{-0.000}$& $-0.44^{+0.004}_{-0.265}$& $-0.49^{+0.004}_{-0.399}$& $-0.30^{+0.000}_{-0.099}$\\
2& $18.3^{+0.44}_{-0.00}$& $-1.1^{+0.00}_{-0.82}$& $-2.1^{+0.58}_{-0.61}$& $-3.3^{+0.18}_{-0.14}$& $-1.06^{+0.165}_{-0.000}$& $-0.04^{+0.000}_{-0.004}$& $-1.06^{+0.072}_{-0.000}$& $-1.30^{+0.000}_{-0.179}$& $-0.64^{+0.157}_{-0.000}$\\
3& $19.2^{+0.27}_{-0.22}$& $-1.0^{+0.69}_{-0.72}$& $-3.4^{+0.13}_{-0.04}$& $-3.4^{+0.14}_{-0.03}$& $-0.48^{+0.102}_{-0.202}$& $-0.07^{+0.018}_{-0.032}$& $-0.48^{+0.093}_{-0.228}$& $-0.55^{+0.150}_{-0.427}$& $-0.31^{+0.061}_{-0.127}$\\
4& $18.9^{+0.37}_{-0.04}$& $-1.2^{+0.69}_{-0.52}$& $-2.8^{+0.50}_{-0.23}$& $-3.2^{+0.15}_{-0.13}$& $-0.74^{+0.233}_{-0.063}$& $-0.06^{+0.005}_{-0.052}$& $-0.74^{+0.204}_{-0.066}$& $-0.97^{+0.179}_{-0.108}$& $-0.43^{+0.131}_{-0.047}$\\
5& $20.0^{+0.00}_{-0.00}$& $-0.4^{+0.00}_{-0.00}$& $-3.4^{+0.00}_{-0.00}$& $-3.4^{+0.00}_{-0.00}$& $-0.17^{+0.000}_{-0.000}$& $-0.09^{+0.000}_{-0.000}$& $-0.18^{+0.000}_{-0.000}$& $0.01^{+0.000}_{-0.000}$& $-0.16^{+0.000}_{-0.000}$\\
Combined& --& --& --& --& $-0.56^{+0.044}_{-0.051}$& $-0.08^{+0.008}_{-0.009}$& $-0.58^{+0.037}_{-0.033}$& $-0.69^{+0.081}_{-0.100}$& $-0.25^{+0.022}_{-0.026}$\\
\hline
\multicolumn{10}{c}{\textbf{J1032+0927}}\\
1& $19.1^{+0.20}_{-0.25}$& $-0.2^{+0.09}_{-0.41}$& $-3.0^{+0.15}_{-0.40}$& $-8.0$& $-0.17^{+0.041}_{-0.213}$& $-0.08^{+0.031}_{-0.063}$& $-0.18^{+0.038}_{-0.231}$& $-0.17^{+0.039}_{-0.270}$& $-0.19^{+0.048}_{-0.069}$\\
2& $19.9^{+0.06}_{-0.02}$& $-0.1^{+0.06}_{-0.03}$& $-2.1^{+0.02}_{-0.06}$& $-8.0$& $-0.04^{+0.015}_{-0.017}$& $-0.03^{+0.015}_{-0.018}$& $-0.04^{+0.015}_{-0.017}$& $-0.02^{+0.014}_{-0.022}$& $-0.06^{+0.015}_{-0.018}$\\
3& $18.6^{+0.63}_{-0.34}$& $-1.0^{+0.20}_{-0.32}$& $-2.6^{+0.95}_{-0.51}$& $-8.0$& $-0.45^{+0.238}_{-0.417}$& $-0.13^{+0.039}_{-0.000}$& $-0.47^{+0.248}_{-0.444}$& $-0.50^{+0.295}_{-0.554}$& $-0.38^{+0.187}_{-0.273}$\\
Combined& --& --& --& --& $-0.07^{+0.013}_{-0.014}$& $-0.03^{+0.013}_{-0.018}$& $-0.06^{+0.017}_{-0.017}$& $-0.08^{+0.022}_{-0.028}$& $-0.09^{+0.017}_{-0.019}$\\
\hline
\multicolumn{10}{c}{\textbf{J2251$-$1227}}\\
1& $18.4^{+0.00}_{-0.00}$& $-2.5^{+0.49}_{-0.00}$& $-1.7^{+0.00}_{-0.36}$& $-8.0$& $-1.22^{+0.005}_{-0.000}$& $0.18^{+0.000}_{-0.243}$& $-1.31^{+0.085}_{-0.000}$& $-1.92^{+0.405}_{-0.000}$& \nodata{}\\
2& $19.5^{+0.06}_{-0.01}$& $-1.8^{+0.03}_{-0.25}$& $-2.7^{+0.09}_{-0.06}$& $-8.0$& $-0.27^{+0.033}_{-0.060}$& $-0.06^{+0.020}_{-0.017}$& $-0.30^{+0.035}_{-0.087}$& $-0.42^{+0.054}_{-0.198}$& \nodata{}\\
Combined& --& --& --& --& $-0.50^{+0.045}_{-0.052}$& $-0.05^{+0.015}_{-0.023}$& $-0.43^{+0.035}_{-0.069}$& $-0.62^{+0.126}_{-0.180}$& \nodata{}\\
\hline

\end{tabular}
\end{center}
\end{table*}

\subsubsection*{Validation of ionization modelling analysis}

To check that our ionization correction method is providing realistic modelling parameters for each component, we applied our method to the abundance pattern of the subDLA towards J1332+0052 \citep{Kislitsyn24}. This particular subDLA is a useful test system as the logN(\HI{}) of each component has been measured directly with high precision \citep{Kislitsyn24}. The goal of this test is to see if our method can reproduce the logN(\HI{}) of the four components of this subDLA using only the logN(Si\ion{ii}) and logN(O\ion{i}) of each component, and total logN(\HI{}) of the subDLA. The same analysis was applied to these four components, using only a \cite{Haardt12} ultraviolet background (i.e.~no AGN component to the radiation) at the redshift of the absorber. The \textsc{cloudy} model grid was extended in metallicity (up to \Zcloudy{}~$=-0.5$ dex) and logN(\HI{}) (down to logN(\HI{})/${\rm cm}^{-2}=16.0$) to ensure the measured quantities of each component \citep{Kislitsyn24} were contained within the grid. We also artificially increased the error on the total logN(\HI{}) of this subDLA to $\pm0.05$ as the acceptance rate of the proposed models was extremely low ($\lll 1$\%) using the quoted error in \cite{Kislitsyn24} of $\pm0.004$. We suspect this is a result of the MCMC sampler not having sufficient precision to propose valid models which satisfy the prior requiring that all four components' logN(\HI{}) must sum to the total value within the measured uncertainty, given that there is nearly a three orders of magnitude difference in logN(\HI{}) between components. Despite larger uncertainties in our predictions ($\approx \pm0.4$ dex) than the measured values reported by \cite{Kislitsyn24} ($\approx \pm0.05$ dex), our analysis was able to reproduce the correct ranked order of both the logN(\HI{}) and \Zcloudy{}\footnote{We note the work by \cite{Kislitsyn24} only have an estimate of the metallicity of the subDLA's components using [O/H], which is expected to have minimal ionization and dust corrections.} for all four components, as well as consistent logN(\HI{}) and \Zcloudy{} for the three strongest components. We point out that the predictions for the two strongest components with logN(\HI{})/${\rm cm}^{-2}\geq18.39$ are closer to the measured values compared to the two components with logN(\HI{})/${\rm cm}^{-2}\leq16.8$. This again is likely a limitation of the MCMC model proposals not having sufficient precision given that logN(\HI{}) in all four components must sum to $19.304\pm0.05$ whilst 99.5\% of the \HI{} is contained within the two strongest components.

We note that there are two assumptions that may not hold for this particular test system: (i) the metallicity of this subDLA is [M/H]~$=-1.71$, and thus assuming an intrinsic [Si/O] that matches the VMP DLA population ([M/H]~$<-2$) may not be valid; and (ii) no dust correction is assumed in our analysis even though the expected dust correction for [Si/O] could be as large as $\approx+0.15$~dex at the metallicity of the subDLA \citep{deCia18}. For assumption (i), we note that [Si/O] in the [M/H]$\approx-1.7$ DLA population appears to be consistent with the VMP DLAs \citep{Cooke15,Berg15,Welsh19}, but note that the typical saturation of the O\ion{i} 1302\AA{} line and increasing dust depletion of Si\ion{ii} in DLAs limits an accurate assessment of this assumption. In order to investigate the assumption on the strength of dust depletion, we re-ran the analysis with an additional $+0.15$~dex added to each component's logN(Si\ion{ii}) as a `maximal' dust correction in each component, and were able to obtain nearly identical results as the no-dust assumption for both the predicted logN(\HI{}) and \Zcloudy{}.  The parameter significantly affected by the inclusion of a dust correction was log($n_{H}$); the median log($n_{H}$) is consistently smaller in all four components by $\approx0.1-0.4$ dex with the addition of the `maximal' dust correction.  We are encouraged that this ionization correction method can be used to model the subDLA components individually.

\section{Discussion}

\subsection{Metallicity and \vninety{}}

Table \ref{tab:v90} provides the total metallicities of the five VMP subDLAs. We include two measurements of metallicity: [M/H] is computed based on the total column density of a given ion \citep[preferrentially O\ion{i}, or Si\ion{ii}; following][]{Rafelski12} and [M/H]$_{\rm IC}$ is the ionization-corrected metallicity obtained from combining \Zcloudy{} of the individual components in our \textsc{Cloudy} modelling.  As we were unable to simultaneously measure the O\ion{i} or Si\ion{ii} column density for the subDLAs toward QSOs J0034$+$1639 and J0529$-$3552, we do not have an [M/H]$_{\rm IC}$ measured for these absorbers. We note that for the three absorbers with measured [M/H]$_{\rm IC}$, the uncorrected [M/H] (all based on O\ion{i}) are consistent (and nearly identical for J1032$+$0927 and J2251$-$1227) to [M/H]$_{\rm IC}$.

In addition to metallicity, Table \ref{tab:v90} includes \vninety{}, a metric typically used as a proxy for the absorber's mass \citep{Prochaska97,Moeller13}. We compute \vninety{} following the procedure outlined in \cite{Ledoux06}. In summary, we preferentially chose metal absorption lines that are neither too saturated nor too weak (see Table \ref{tab:v90} for the metal line used for each subDLA). Practically, this means that the selected metal lines have the strongest absorption component between $10-60$ per cent of the continuum flux level. We perform the \vninety{} computation on 1000 realizations of the fitted profiles with random Gaussian noise inserted based on the error spectrum. The mean and standard deviation of these 1000 realizations are provided in Table \ref{tab:v90}. Typically, low-ionization lines are used for the computation of \vninety{}, but since J0034+1639 only exhibits high-ionisation lines, we provide two estimates of \vninety{} using low and high ionisation lines. For the four VMP subDLAs with detected low ionisation lines, we find their \vninety{} measurements are consistent within the scatter of DLAs within the same metallicity range \citep[{[M/H]}~$\lesssim-2$][]{Ledoux06,Neeleman13}. The exception to the apparent \vninety{}--[M/H] relation is J0415$-$4357, which has a much higher \vninety{} owing to the large separation of the three component groups in the low ionization lines. Individually, these three component groups have $20<$~\vninety{}~$<50$~\kms{}, which are consistent with other VMP DLAs. All subDLAs of this sample are consistent with the broad range in \vninety{} found in \zabs{}~$\gtrsim4.5$ absorbers \citep{Poudel20}.

\begin{table*}
\begin{center}
\caption{Metallicity and \vninety{} measurements of the VMP subDLAs}
\label{tab:v90}
\begin{tabular}{lccccc}
\hline
QSO& $z_{\rm abs}$& [M/H] (ion) & [M/H]$_{\rm IC}$ & \vninety{} low ion (line) & \vninety{} high ion (line)\\
& & & & \kms{}& \kms{}\\
\hline
J0034+1639& 4.22636& $<-2.82$ (Si\ion{ii}) & \nodata{}& \nodata{} & $321.6\pm2.1$ (C\ion{iv} 1550)\\
J0415$-$4357& 4.03510& $-2.97\pm0.31$ (O\ion{i}) & $-2.87^{+0.08}_{-0.11}$ &$152.7\pm3.6$ (Si\ion{ii} 1260) & $255.7\pm3.3$ (C\ion{iv} 1550)\\
J0529-3552& 3.68422& $-1.93\pm0.20$ (Si\ion{ii}) & \nodata{} &$16.4\pm2.0$ (Al\ion{ii} 1670)& $196.6\pm2.9$ (Si\ion{iv} 1393)\\
J1032+0927& 3.80385& $-2.02\pm0.16$ (O\ion{i}) & $-2.01^{+0.07}_{-0.08}$ &$20.1\pm6.1$ (Si\ion{ii} 1526)& $92.1\pm11.0$ (C\ion{iv} 1548)\\
J2251-1227& 3.98862& $-2.59\pm0.16$ (O\ion{i}) & $-2.59^{+0.08}_{-0.08}$ &$20.8\pm4.9$ (C\ion{ii} 1334)& $47.7\pm10.1$ (C\ion{iv} 1548)\\
\hline
\end{tabular}
\end{center}
\end{table*}

\subsection{Total abundance ratios}
\label{sec:tot}

\begin{figure*}
\begin{center}
\includegraphics[width=\textwidth]{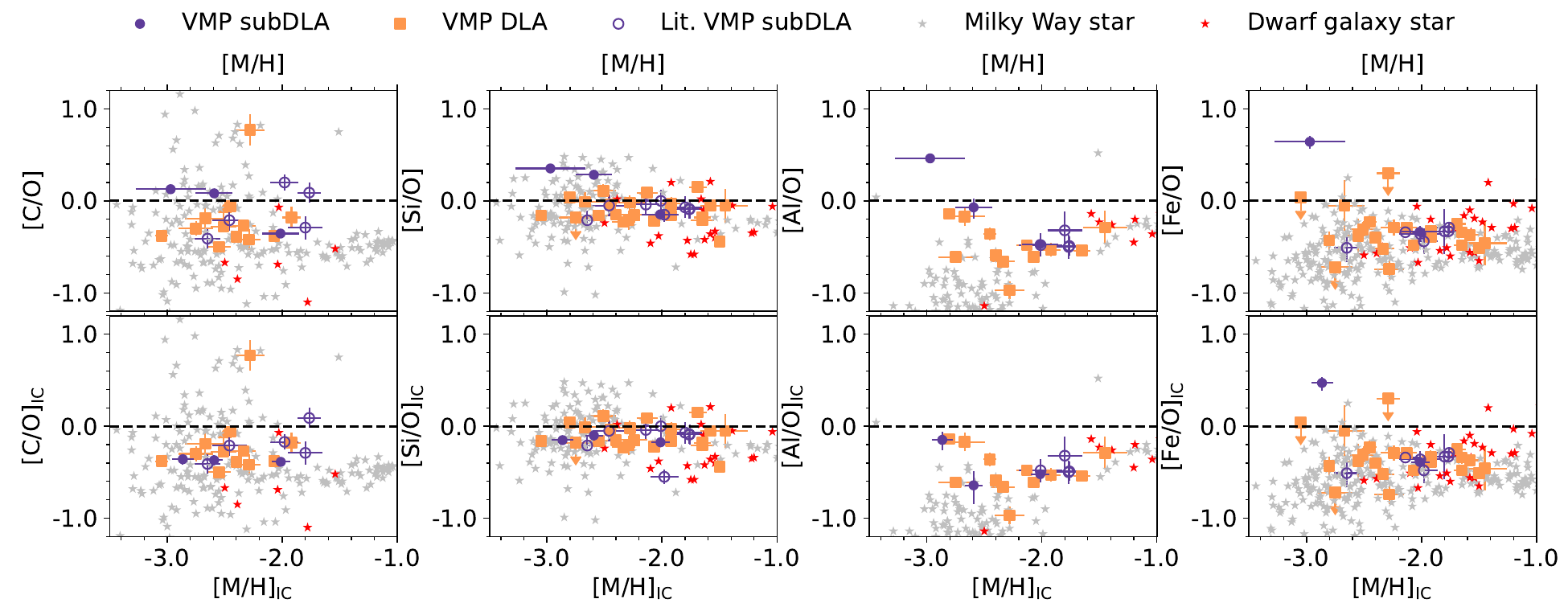}
\caption[Abundance ratios of subDLAs and VMP DLAs]{Top panels -- The measured abundance ratios \ensuremath{[\mathrm{C}/\mathrm{O}]}, \ensuremath{[\mathrm{Si}/\mathrm{O}]}, \ensuremath{[\mathrm{Al}/\mathrm{O}]}, and \ensuremath{[\mathrm{Fe}/\mathrm{O}]} (from left to right) as a function of metallicity (\ensuremath{[\mathrm{M}/\mathrm{H}]}) for subDLAs (purple circles) and literature VMP DLAs and subDLAs \citep[orange squares and hollow purple circles][]{Cooke17,Morrison16,Welsh19}. Stellar literature abundances are denoted as stars, with grey stars representing Milky Way measurements \citep{Abohalima18,Amarsi19,Jofre15,Nordlander17,Prakapavicius17}, and red stars for Local Group dwarf galaxy abundances \citep{Abohalima18}. Apart from the literature VMP subDLA from \citet{Morrison16}, ionization corrections (ICs) were not included in the literature abundance ratios. An additional error of \ensuremath{\pm 0.1} dex has been included for the literature VMP subDLA measurements to account for the typical range of ICs estimated for these systems \citep[e.g.][]{DessaugesZavadsky03,Pettini08,Cooke11}. \ensuremath{[\mathrm{M}/\mathrm{H}]} is taken as \ensuremath{[\mathrm{O}/\mathrm{H}]} (preferentially) or \ensuremath{[\mathrm{Si}/\mathrm{H}]} for the subDLAs (Table~\ref{tab:v90}) and DLAs, while \ensuremath{[\mathrm{Fe}/\mathrm{H}]} is used for stellar metallicities. Note that the error in the abundance ratio \ensuremath{[\mathrm{X}/\mathrm{O}]} is typically smaller than the points themselves. Bottom panels -- The same as the top panels, but showing the subDLA abundance ratios including ICs (\ensuremath{[\mathrm{X}/\mathrm{O}]_{\mathrm{IC}}}). The horizontal black dashed line represents the solar relative abundance.}

\label{fig:XO}
\end{center}
\end{figure*}

The top panels of Figure \ref{fig:XO} show the abundance ratios [C/O], [Si/O], [Al/O], and [Fe/O] using the total column densities for the subDLAs presented in Table \ref{tab:Ntot} (purple circles). For comparison, we also include the compilation of 22 VMP subDLAs (open purple circles) and DLAs (orange squares) from \cite{Cooke17}, \cite{Morrison16} and \cite{Welsh19} which have been observed with equivalent high spectral resolution in order to provide accurate column density measurements \citep[e.g.~see Sec.~7.3 of][]{Cooke11}. The bottom panels of Figure \ref{fig:XO} show the same abundance ratios, but with the ICs included for the subDLAs (Table \ref{tab:ICX}\footnote{We emphasize that the uncertainties on the abundances in Table \ref{tab:ICX} are determined based on the posterior distribution from the ionization correction modelling. The 0.15--0.3 dex uncertainties on logN(\HI{}) are included on [X/H] but not [X/O] as the [X/O] abundance ratio measurement is independent of logN(\HI{}).}). Upon inclusion of the ICs, the subDLAs in this work appear consistent with the VMP DLA abundance pattern. However, we note that [Fe/O] and [Al/O] in the subDLA towards J0415$-$4357 are on the high end of the abundance pattern of the VMP DLA population, but could be more in line with extremely metal-poor DLAs \citep{Welsh24}. In comparison to the previous study of [M/H]~$\approx-2$ subDLAs by \cite{Morrison16}, our subDLA abundance pattern is consistent with their abundance pattern prior to ICs but not with ICs included. We suspect the discrepancy is the result of the different methods adopted to determine the ICs, as \cite{Morrison16} used the C\ion{ii}/C\ion{iv} and Si\ion{ii}/Si\ion{iv} column density ratios to determine the ICs. Such an assumption implies that the low and high ionization state gas is co-spatial, which is likely not the case (e.g.~comparing components for low and high ionization gas in Tables \ref{tab:loComp} and \ref{tab:hiComp}). 

For comparison, we also show the Galactic and Local Group dwarf galaxy stellar abundances from several literature compilations \citep[including non-local thermodynamic equilibrium corrections, NLTE\footnote{It is unclear how uniform NLTE corrections are applied to all elemental abundances for a given star in the \cite{Abohalima18} compilation. While we require the `NLTE' flag in this compilation, we caution not all abundance may have NLTE corrections applied. }][]{Abohalima18,Amarsi19,Jofre15,Nordlander17,Prakapavicius17}. As previously seen for VMP DLAs \citep[e.g.][]{Cooke11}, [C/O] in the VMP subDLAs also shows an elevated value compared to stars \citep[in particular when compared to the NLTE-corrected abundances from][or stars with $\rm{[Fe/H]}$~$>-2.5$]{Amarsi19}. While the near agreement of [Si/O] between the VMP subDLAs and VMP DLAs is a result of our assumptions in the IC modelling, it is encouraging that the VMP DLAs also agree with the observed [Si/O] in both Milky Way and dwarf galaxy stars. This result is consistent with \cite{Saccardi23}, where they found more diffuse absorption systems (i.e.~Lyman limit systems, including subDLAs) have similar abundance patterns to very metal poor stars in the Local Group. The most obvious discrepancy between stars and VMP DLAs and subDLAs appears with the observed [Al/O]. The large scatter at [M/H]~$\lesssim-2.5$ in the absorber population could be explained by stochastic enrichment resulting from different stellar masses or supernovae types \citep[][]{Heger02, HegerWoosley2010}. However for stars, it is difficult to determine if the scatter is also driven by variations in the NLTE corrections as both Al and O are very sensitive to NLTE effects \citep[e.g.][]{Amarsi19,Lind24,Nordlander17}. Unfortunately, very few measurements of [Al/O] with NLTE corrections exists in [M/H]~$\leq-2.5$ stars to enable a more accurate comparison.

\begin{table*}

\begin{center}
\caption{Ionization-corrected abundances  and [X/O] abundance ratios of the VMP subDLAs}
\label{tab:ICX}
\resizebox{\textwidth}{!}{
\begin{tabular}{lcccccccccc}

\hline
QSO& \zabs{}& [O/H]\subIC{}& [C/H]\subIC{}& [C/O]\subIC{}& [Si/H]\subIC{}& [Si/O]\subIC{}& [Al/H]\subIC{}& [Al/O]\subIC{}& [Fe/H]\subIC{}& [Fe/O]\subIC{}\\
\hline
J0415$-$4357& 4.03510& $-3.05_{-0.31}^{+0.31}$& $-3.41_{-0.31}^{+0.31}$& $-0.36_{-0.06}^{+0.05}$& $-3.20_{-0.30}^{+0.30}$& $-0.15_{-0.04}^{+0.04}$& $-3.20_{-0.32}^{+0.31}$& $-0.15_{-0.11}^{+0.09}$& $-2.58_{-0.31}^{+0.31}$& $0.47_{-0.08}^{+0.07}$\\
J1032+0927& 3.80385& $-2.05_{-0.16}^{+0.16}$& $-2.44_{-0.16}^{+0.16}$& $-0.39_{-0.03}^{+0.03}$& $-2.23_{-0.15}^{+0.15}$& $-0.17_{-0.03}^{+0.03}$& $-2.57_{-0.15}^{+0.15}$& $-0.52_{-0.04}^{+0.04}$& $-2.44_{-0.16}^{+0.15}$& $-0.39_{-0.05}^{+0.04}$\\
J2251$-$1227& 3.98862& $-2.64_{-0.16}^{+0.16}$& $-3.01_{-0.17}^{+0.17}$& $-0.37_{-0.06}^{+0.06}$& $-2.74_{-0.17}^{+0.15}$& $-0.10_{-0.05}^{+0.04}$& $-3.29_{-0.26}^{+0.22}$& $-0.64_{-0.20}^{+0.15}$& \nodata{}& \nodata{}\\
\end{tabular}
}
\end{center}
\end{table*}

\subsection{Component abundance ratios}
\label{sec:compAbund}

\begin{figure}
\begin{center}
\includegraphics[width=0.5\textwidth]{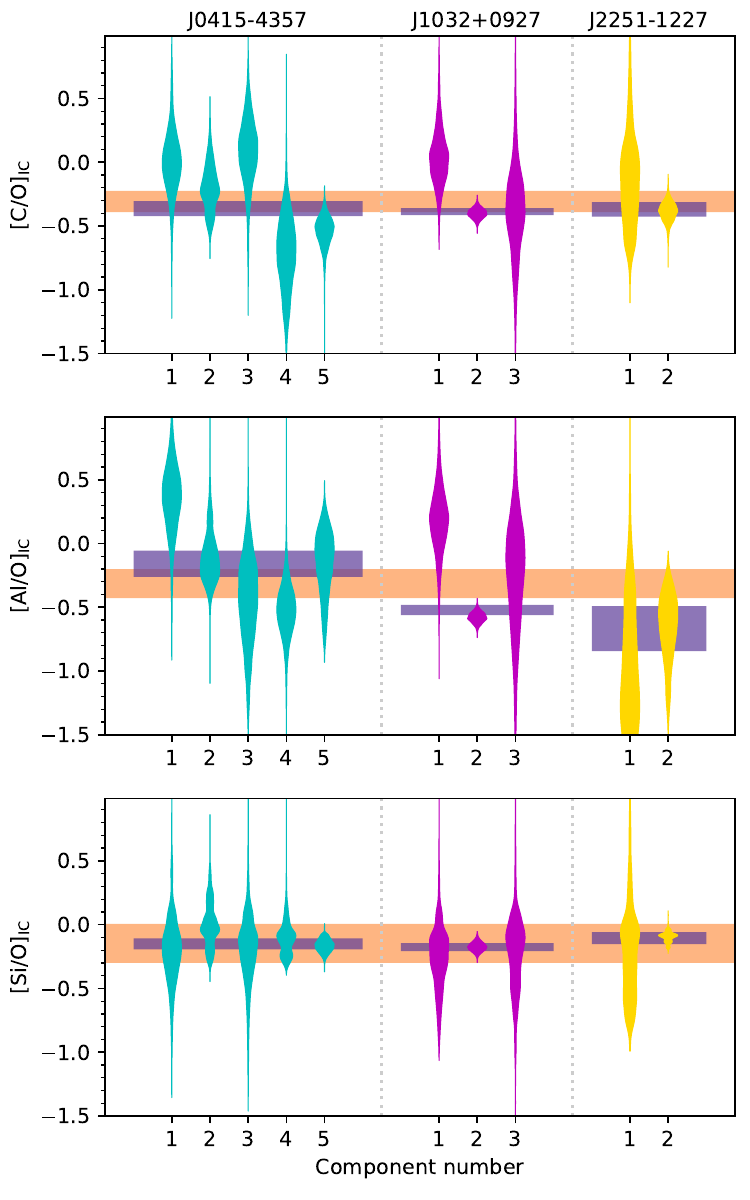}
\caption{Each panel shows a violin plot of the ionization-corrected abundance ratios [C/O], [Si/O], and [Al/O] of each component for the three subDLAs in which we were able to obtain ICs (J0415$-$4357, J1032+0927, and J2251$-$1227; cyan, magenta, and yellow, respectively). The components of each subDLA are separated by the vertical dotted grey line. The widths of each violin denotes the probability density of a given abundance ratio in that component. The component numbers are the same as in Table \ref{tab:loComp}.  The horizontal orange band in each panel represents the mean and error in the respective abundance ratio for the VMP DLA literature sample \citep{Cooke15,Welsh19}. The consistency in [Si/O]$_{\rm IC}$ between components and the VMP DLA orange band is a result of the method for determining the IC. For reference, the horizontal purple bars behind each set of violins denote the total abundance ratio and error for the entire subDLA (Table \ref{tab:ICX}), and highlights how discrepant each component is from the total of the subDLA. There are several pairs of components within a given subDLA where the [C/O] or [Al/O] are discrepant at $\geq 2\sigma$ significance (e.g.~components 3 and 5 in J0415$-$4357 for [C/O]; components 1 and 2 in J1032+0927 for [C/O] and [Al/O]) suggesting chemical inhomogeneities between components in these two subDLAs.}
\label{fig:XOcomps}
\end{center}
\end{figure}

We now take advantage of our component level \cloudy{} modelling to explore the uniformity and consistency of the metal abundances among the absorption components within a given subDLA. We first remark that dust depletion is minimal in metal-poor systems \citep{deCia18}, and would not show such significant variations between neighbouring clouds \citep{Vladilo11,Vladilo18}. Figure \ref{fig:XOcomps} shows violin plots of the ionization-corrected abundance ratios [C/O]\subIC{}, [Al/O]\subIC{}, and [Si/O]\subIC{} for the three subDLAs where we were able to derive ICs. The distribution of the abundance ratios, which defines the violins' shapes, is obtained by using the same resampling of the \textsc{alis} covariance matrix used to compute the column densities and ICs (Sections \ref{sec:cols} and \ref{sec:Ion}). We point out that there are several components where the column densities are measured at $\leq2\sigma$ significance (Table \ref{tab:loComp}), which leads to the long extensions of the violins beyond $\pm1.0$~dex. The width of these extended tails thus can be interpreted as the likelihood that the abundance ratio of this component is an upper ($\lesssim-1.0$ dex) or lower limit ($\gtrsim+1.0$ dex). The reference average VMP DLA abundance ratios from the literature sample are denoted by the horizontal orange bands in Figure \ref{fig:XOcomps}. The agreement between the VMP DLAs and all components of the subDLAs for [Si/O] (bottom panel of Figure \ref{fig:XOcomps}) is a result of our IC method of fixing [Si/O] to within the orange band. Furthermore, whilst there is generally good agreement between the total abundance ratio of the individual subDLAs to the VMP DLA distribution, it is clear that the dispersion in the component abundance ratios is much greater than the VMP DLA spread.

In order to search for potential discrepancies in the abundance ratios [C/O] and [Al/O] amongst the absorption components, we computed $\Delta$[X/O]$_{\rm IC}$ -- the relative difference in the abundance ratio between two components -- using pairs of abundance ratios from the same sampling of the covariance matrix. Table \ref{tab:compdXO} shows the mean $\Delta$[X/O]$_{\rm IC}$ between every combination of two components for each subDLA, computed relative to [X/O] of the second component. $\Delta$[Si/O]$_{\rm IC}$ is included for completeness. The significance of the measured $\Delta$[X/O]$_{\rm IC}$ from 0 (denoted in square brackets in Table \ref{tab:compdXO}) is computed as the mean $\Delta$[X/O]$_{\rm IC}$ divided by the standard deviation of $\Delta$[X/O]$_{\rm IC}$. The boldface entries in Table \ref{tab:compdXO} identify where $\Delta$[X/O]$_{\rm IC}$ is measured to have a  $\geq2\sigma$ significant discrepancy between components.

There are several notable components in the subDLA towards J0415$-$4357 which suggest strong $\Delta$[X/O]$_{\rm IC} \geq0.5$ dex. Component 1 consistently has a higher [Al/O] compared to other components (at 1.3--2.3$\sigma$ significance) whilst component 3 has a higher [C/O] compared to other components (0.2--2.3$\sigma$). However both components 1 and 3 are quite weak (Figure \ref{fig:J0415vel}), and can be sensitive to the continuum placement and noise within the observed spectrum.  Components 4 and 5 also appears to have systematically lower [C/O] compared to the other components, in particular relative to the strongly detected component 2 ($\approx1.5\sigma$ significance). Additionally, there is a $>2.2\sigma$ discrepancy in both [C/O] and [Al/O] between components 1 and 2 in the subDLA towards J1032+0927.

These discrepancies highlight the possibility that there are inhomogeneities in the abundance patterns between components within the same subDLA absorber. Based on this analysis and the assumptions made in our ionization modelling (Section \ref{sec:Ion}), these observed inhomogeneities can be explained by at least one of the following reasons:

\begin{itemize}
    \item[1.] The assumptions made in constructing our \cloudy{} ionization modelling grid (including geometry and ionizing source) are incorrect. While we currently cannot confirm the physical conditions and geometry of the gas in these subDLAs, we note that the assumptions made are quite typical of ICs derived in the literature \citep[e.g.][]{Milutinovic10,Morrison16,Fumagalli16b}. Modifications to these assumptions would require the community to revisit our general implementation of \cloudy{} modelling of absorption line systems.
    
    \item[2.] Our assumption that [Si/O]\subIC{} in all subDLA components must be identical, used for the ionization modelling, is incorrect. If there is indeed variations in the intrinsic [Si/O] of each component, then naturally there are inhomogeneities in [Si/O] between subDLA components prior to the ionization modelling. Furthermore, we note that the DLA abundance ratio assumed for the subDLA components (${\rm [Si/O]_{DLA}=-0.15\pm0.15}$) is consistent with the typical value seen in metal-poor stars for both the Milky Way and local dwarf galaxies (Figure \ref{fig:XO}). Assuming a different intrinsic [Si/O] would also imply a different abundance pattern not typically seen in metal-poor environments. 

    \item[3.] There are indeed chemical inhomogeneities in these two VMP subDLAs, which can be caused by inhomogeneous mixing of the gas or or stochasticity in the nucleosynthetic channels.
\end{itemize}

In summary, two of these three reasons imply that there are indeed intrinsic inhomogeneities between the components of the same absorber.

While we caution on the significance of the observed $\Delta$[C/O]$_{\rm IC}$ and $\Delta$[Al/O]$_{\rm IC}$ discrepancies in the subDLA towards J0415$-$4357, we note that the well-separated velocity components 2, 4, and 5 (see Figure \ref{fig:J0415vel}) are likely to not have mixed given their $\approx60$~\kms{} separations and low Doppler parameters (Table \ref{tab:loComp}). This suggests the subDLA components show an intrinsic scatter. A possible explanation of this scatter is that the subDLA was enriched via a stochastic process driven by a handful of stars. We will explore this possibility further in Section \ref{sec:model}.

\begin{table*}
\begin{center}
\caption{Relative differences in abundance ratios ($\Delta$[X/O]$_{\rm IC}$) between components. Bold entries denote differences at $\geq2\sigma$ signficance.}
\label{tab:compdXO}
\begin{tabular}{lcccc}
\hline
1$^{\rm st}$ Component& 2$^{\rm nd}$ Component& $\Delta$[C/O]$_{\rm IC}$& $\Delta$[Al/O]$_{\rm IC}$& $\Delta$[Si/O]$_{\rm IC}$\\
\hline
\multicolumn{5}{c}{\textbf{J0415$-$4357}}\\
1& 2& $0.20~[0.6\sigma]$& $0.48~[1.3\sigma]$& $-0.20~[0.6\sigma]$\\
1& 3& $-0.08~[0.2\sigma]$& $0.82~[1.8\sigma]$& $-0.01~[0.0\sigma]$\\
1& 4& $0.71~[1.8\sigma]$& $\mathbf{0.89~[2.3\sigma]}$& $-0.11~[0.3\sigma]$\\
1& 5& $0.55~[1.9\sigma]$& $0.52~[1.3\sigma]$& $-0.04~[0.1\sigma]$\\
2& 3& $-0.28~[0.9\sigma]$& $0.34~[0.8\sigma]$& $0.19~[0.5\sigma]$\\
2& 4& $0.51~[1.5\sigma]$& $0.41~[1.3\sigma]$& $0.09~[0.4\sigma]$\\
2& 5& $0.35~[1.6\sigma]$& $0.04~[0.1\sigma]$& $0.17~[0.9\sigma]$\\
3& 4& $0.80~[2.0\sigma]$& $0.07~[0.2\sigma]$& $-0.10~[0.3\sigma]$\\
3& 5& $\mathbf{0.63~[2.1\sigma]}$& $-0.30~[0.7\sigma]$& $-0.02~[0.1\sigma]$\\
4& 5& $-0.17~[0.5\sigma]$& $-0.37~[1.1\sigma]$& $0.07~[0.4\sigma]$\\
\multicolumn{5}{c}{\textbf{J1032+0927}}\\
1& 2& $0.45~[2.0\sigma]$& $\mathbf{0.77~[2.7\sigma]}$& $-0.09~[0.3\sigma]$\\
1& 3& $0.42~[1.0\sigma]$& $0.44~[0.8\sigma]$& $-0.05~[0.1\sigma]$\\
2& 3& $-0.02~[0.1\sigma]$& $-0.33~[0.7\sigma]$& $0.04~[0.1\sigma]$\\
\multicolumn{5}{c}{\textbf{J2251$-$1227}}\\
1& 2& $0.25~[0.7\sigma]$& $-0.32~[0.5\sigma]$& $-0.09~[0.2\sigma]$\\
\hline

\end{tabular}
\end{center}
\end{table*}

\subsection{Chemical enrichment modelling}
\label{sec:model}

The observed scatter across the abundance patterns of metal-poor absorption line systems has been shown to be an invaluable tool to investigate their chemical enrichment histories \citep{Welsh19, Welsh24, Vanni23, Sodini24}.  In these examples, the scatter is calculated between the different absorbers in the sample. The three systems presented in this work provide the unique opportunity to investigate the source of scatter within a single absorption line system. In order to investigate whether these differences in components' abundance ratios can be explained by different chemical enrichment events, we modelled the expected nucleosynthetic yields from different metal-poor and pristine stellar progenitors and matched them to the observations in order to search for different enrichment events between components.  Distinguishing between Population \textsc{ii} and Population \textsc{iii} supernovae yields is difficult due to degeneracies in the abundance ratio signatures \citep[e.g.][]{Cooke14,Welsh24}.  For the abundances considered in this work, the yields from both Population \textsc{ii} and Population \textsc{iii} core-collapse supernovae are broadly consistent \citep{Welsh19,Vanni24}. This can be seen through a comparison of the \cite{Woosley95} and \cite{HegerWoosley2010} yields. Despite this degeneracy, the signature of Population \textsc{iii} stars can still be teased out in a number of ways. In particular, Population \textsc{iii} stars are expected to form in small multiples. This can lead to a large amount of scatter in the observed abundance patterns of Population \textsc{iii} enriched systems. Previous work, employing a stochastic chemical enrichment model, has indicated that VMP DLAs have undergone few enrichment events. VMP DLAs with similar metallicites to those presented in this work appear to be enriched by $\sim 10$ SNe \citep[with a 95 per cent upper limit of 72 SNe;][]{Welsh19}. We note that we cannot distinguish whether or not these SNe are predominantly Population \textsc{ii} or Population \textsc{iii}. We use the scatter observed within these individual subDLA components to investigate the properties of the individual progenitors that may be responsible for enriching metal-poor gas at $z\sim 3-4$. We utilise the core collapse supernovae yields from \citep{HegerWoosley2010} to represent both Population \textsc{ii} and Population \textsc{iii} progenitors owing to the extensive range of progenitor properties that were explored during their calculations. The \citet{HegerWoosley2010} simulations trace nucleosynthesis processes within stellar progenitors across their lifetimes and calculate the ejected yields of elements following the eventual core-collapse supernovae of these stars. These simulations explore initial progenitor masses in the range $M=(10-100)~{\rm M_{\odot}} $, explosion energies from $E_{\rm exp}=(0.3-10)\times10^{51}$ erg, and mixing parameters from $f_{\rm He} = 0 - 0.25$. The explosion energy is a measure of the final kinetic energy of the ejecta at infinity, while the mixing between stellar layers is parameterised as a fraction of the helium core size. This parameter space is evaluated across 120 masses, 10 explosion energies, and 14 mixing parameters. We consider the entire discrete parameter space up to a progenitor mass of  $M = 70 {\rm M}_{\odot}$. This corresponds to the transition at which pulsational pair instability supernovae are thought to occur \citep{Woosley17}.

We conduct a brute force maximum likelihood analysis to determine the posterior distribution of each model parameter (i.e. progenitor mass, explosion energy, and mixing parameter) for each absorption component by using the ionization-corrected abundance distributions per component described in Section \ref{sec:compAbund}. We only consider the [C/O], [Al/O], and [Si/O] abundance ratios, and evaluate the model parameter likelihood for each of the 10\,000 simulated combinations of abundance ratios in linear space. In this scenario, we treat the associated errors of each combination as negligible such that the underlying errors are captured in the resulting distributions of progenitor properties from the 10\,000 maximum likelihood evaluations. 

\begin{figure*}
    \begin{center}
\includegraphics[width=\textwidth]{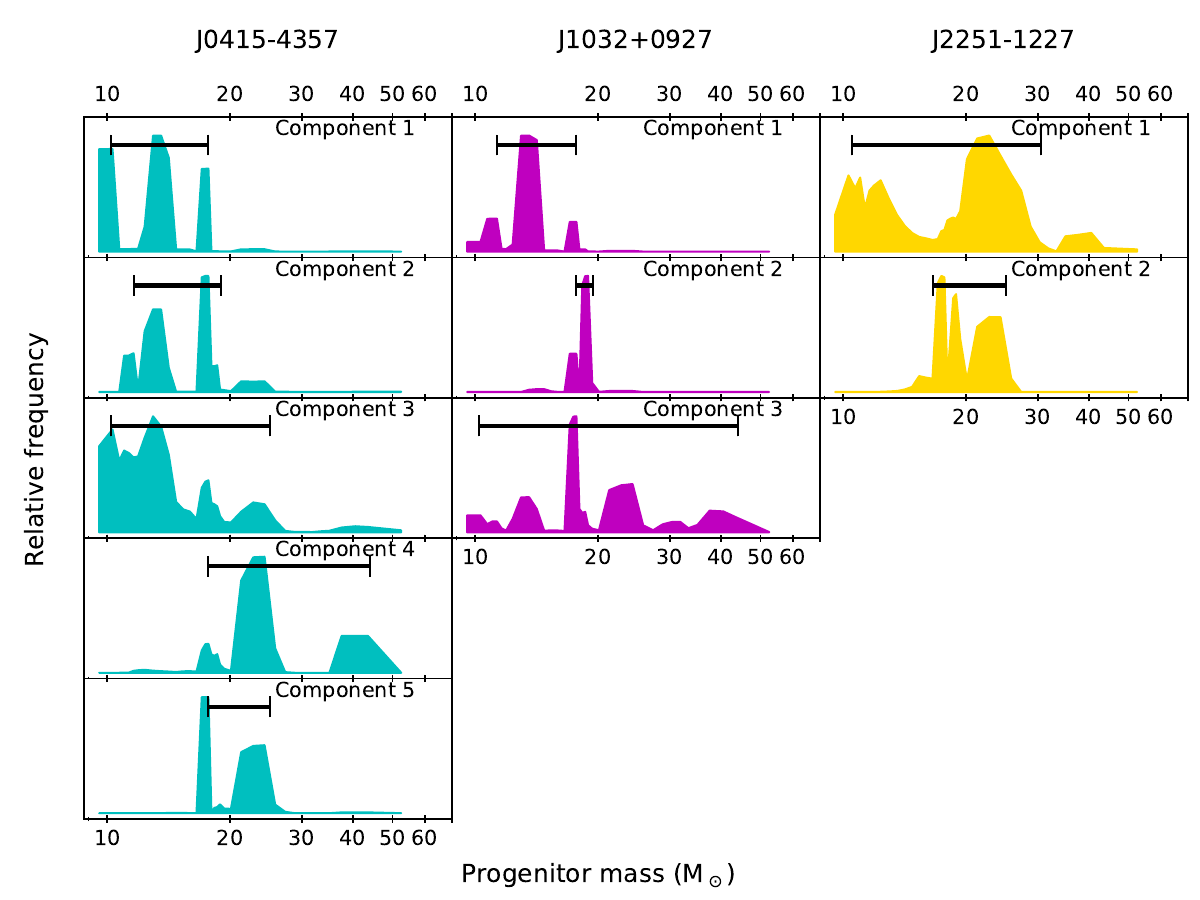}
\caption{The posterior distribution of the modelled progenitor mass for each component of the three subDLAs towards J0415$-$4357 (left column), J1032+0927 (middle), and J2251$-$1227 (right). The distributions have been smoothed using a top-hat kernel with a width of three bins. The horizontal black errorbars denote the 90\% confidence intervals in the progenitor mass estimate for each component. For all components in all three subDLAs, the favoured progenitor mass is $\lesssim30$~\Msol, but there appears to be differences in the distributions between components in both subDLAs towards J0415$-$4357 (components 2 and 4 or 5) and J1032+0927 (components 1 and 2).}
\label{fig:masshists}
\end{center}
\end{figure*}

The resulting progenitor mass posterior distributions for each component of the three subDLAs are shown in Figure \ref{fig:masshists}. A result from this modelling is that all components for the three subDLAs favour a median progenitor mass of $10\lesssim{\rm M}\lesssim30$~\Msol. This is similar to the mass range probed by the VMP DLA population \citep{Welsh19}. This result is likely in part due to the similarity in the underlying abundance pattern and thus enrichment history between the two types of absorbers. If the enrichment is a result of Population \textsc{iii} stellar enrichment, this could be the signature of a shallow (i.e.~top light) initial mass function of the first generation stars, or that more massive stars do not contribute nucleosynthetically due to direct collapse into a black hole \citep{Heger03,Sukhbold16}. This preference towards low mass enriching progenitors is also seen when analysing the chemistry of metal-poor stars in the Milky Way \citep{Ishigaki2018} and $z>6$ intervening quasars absorbers \citep{Christensen23}. Focussing on the noted individual components at the end of Section \ref{sec:compAbund}, we also recover  subtle differences in the progenitor mass distributions between component 2 (with a $2\sigma$ confidence interval of $11.6\leq {\rm M} \leq 19.0$~\Msol{}) and 4 ($17.7\leq {\rm M} \leq 44.0$~\Msol{}) or 5 in J0415$-$4357 ($17.7\leq {\rm M} \leq 25.0$~\Msol{}), as well as for components 1 ($11.3\leq {\rm M} \leq 17.7$~\Msol{}) and 2 ($17.7\leq {\rm M} \leq 19.4$~\Msol{}) in J1032+0927. The posterior distributions of explosion energy and mixing parameters are provided for reference in the appendix (Figures \ref{fig:exphists} and \ref{fig:mixhists}, respectively). We note that the explosion energy and mixing parameters are generally not sufficiently constrained by the data to pick up differences between components. The exception is the $\approx2\sigma$ discrepancy in the predicted progenitor explosion energy for components 1 ($2\sigma$ confidence interval of $\approx0.6\times10^{51} - 2\times10^{51}$~erg) and 2 ($\approx2\times10^{51} - 5\times10^{51}$~erg) for the subDLA towards J1032+0927.

These results do suggest the possibility that the most discrepant observed variations in the abundance ratios between components within two of our subDLAs can be reproduced by different enrichment events, and that the gas between these enrichment events has not yet mixed into the various components. This is inline with recent work that highlights the most metal-poor DLAs at $ 2<z<3.7$ may have been enriched by only a handful of core-collapse supernovae \citep{Welsh24}. The subDLAs studied here are found at earlier epochs meaning the ejecta from these supernovae have had less time to mix. Applying this component by component analysis to both subDLAs and DLAs may therefore offer the unique opportunity to determine the properties of the individual progenitors contributing to the enrichment of very metal-poor gas.

\section{Conclusion}

In this paper we investigated the chemical abundance ratios [C/O], [Si/O], [Al/O], and [Fe/O] of five VMP subDLAs at \zabs{}~$\approx4$, and compared their values to VMP DLAs and Local Group stellar abundances in the literature. In order to account for ionization effects, we developed a novel technique to determine an ionization correction (IC) on a per-component basis using \cloudy{}. By requiring the observed [Si/O] of each subDLA component to match the [Si/O] distribution of VMP DLAs, we infer the most likely \cloudy{} model parameters, thereby allowing us to calculate ICs for every component of all detected ions (Section \ref{sec:Ion}). Upon correcting each subDLA component for ionization effects, we find:

\begin{itemize}
\item[1.] Metal poor subDLAs and  VMP DLAs appear to have similar [C/O], [Al/O], and [Fe/O] (Figure \ref{fig:XO}), and likely trace similar nucleosynthetic environments. Nucleosynthetic modelling of three of the subDLAs suggests that yields of metal-poor stars with progenitor masses of $\lesssim30$~\Msol\ can reproduce the observed abundances of these metal-poor subDLAs.

\item[2.] In three of the subDLAs in our sample, we find tentative evidence of inhomogeneities in [C/O] and [Al/O] between components of the same systems (Figure \ref{fig:XOcomps}). These inhomogeneities can be reproduced by different progenitor masses  or supernovae explosion energies (Figures \ref{fig:masshists} and \ref{fig:exphists}), suggesting isolated enrichment events may have occurred and the gas has yet to mix.

\end{itemize}

All of these results suggest that VMP subDLAs are ideal targets to study metal-poor gaseous reservoirs \citep[e.g.][]{Saccardi23}, and are highly complementary to VMP DLA studies. As an added bonus, subDLAs have a higher incidence rate than DLAs, and the lines of VMP subDLAs are rarely saturated. During the forthcoming era of 40m-class telescopes with spectrographs like the Armazones High Dispersion Echelle Spectrograph \citep[ANDES][]{Dodorico23}, improved precision on detailed abundance measurements in individual components will be key to study the yields of the first stars. In this work we have demonstrated that a component-by-component comparison can offer a unique insight into the mixing properties of near-pristine gas.

\section*{Acknowledgments}
We thank Jason X. Prochaska for useful discussions, Michael Murphy for his support for using  UVES\_\textsc{popler}, and Kim Venn for her suggestions on the databases used for the stellar comparison literature sample. We are also grateful to the anonymous referee for their time providing feedback on this manuscript. During this work, RJC was funded by a Royal Society University Research Fellowship. RJC acknowledges support from STFC (ST/T000244/1, ST/X001075/1). LC is supported by DFF/Independent Research Fund Denmark, grant-ID 2032–00071. SL acknowledges support by FONDECYT grant 1231187. The results of this manuscript are based on observations collected at the European Organisation for Astronomical Research in the Southern Hemisphere under ESO programmes 094.A-0233(A), 0104.A-0149(A), and 110.23QP.

\section*{Data Availability}
The data underlying this article are available on the European Southern Observatory archive facility at \url{https://archive.eso.org/eso/eso_archive_main.html}, and can be found by searching for the program identification numbers 094.A-0223(A), 0104.A-0149(A), and 110.23QP.

\bibliography{bibref}

\begin{thebibliography}{93}
\expandafter\ifx\csname natexlab\endcsname\relax\def\natexlab#1{#1}\fi

\bibitem[{{Abohalima} \& {Frebel}(2018)}]{Abohalima18}
{Abohalima} A., {Frebel} A., 2018, \apjs, 238, 36

\bibitem[{{Amarsi} {et~al}\mbox{.}(2019){Amarsi}, {Nissen}, \&
  {Sk{\'u}lad{\'o}ttir}}]{Amarsi19}
{Amarsi} A.~M., {Nissen} P.~E., {Sk{\'u}lad{\'o}ttir} {\'A}., 2019, \aap, 630,
  A104

\bibitem[{{Asplund} {et~al}\mbox{.}(2009){Asplund}, {Grevesse}, {Sauval}, \&
  {Scott}}]{Asplund09}
{Asplund} M., {Grevesse} N., {Sauval} A.~J., {Scott} P., 2009, \araa, 47, 481

\bibitem[{{Beers} {et~al}\mbox{.}(1992){Beers}, {Preston}, \&
  {Shectman}}]{Beers1992}
{Beers} T.~C., {Preston} G.~W., {Shectman} S.~A., 1992, \aj, 103, 1987

\bibitem[{{Berg} {et~al}\mbox{.}(2015){Berg}, {Ellison}, {Prochaska}, {Venn},
  \& {Dessauges-Zavadsky}}]{Berg15}
{Berg} T.~A.~M., {Ellison} S.~L., {Prochaska} J.~X., {Venn} K.~A.,
  {Dessauges-Zavadsky} M., 2015, \mnras, 452, 4326

\bibitem[{{Berg} {et~al}\mbox{.}(2019){Berg}, {Ellison},
  {S{\'a}nchez-Ram{\'\i}rez}, {L{\'o}pez}, {D'Odorico}, {Becker},
  {Christensen}, {Cupani}, {Denney}, \& {Worseck}}]{Berg19}
{Berg} T. A.~M. {et~al.}, 2019, \mnras, 488, 4356

\bibitem[{{Berg} {et~al}\mbox{.}(2016){Berg}, {Ellison},
  {S{\'a}nchez-Ram{\'{\i}}rez}, {Prochaska}, {Lopez}, {D'Odorico}, {Becker},
  {Christensen}, {Cupani}, {Denney}, \& {Worseck}}]{Berg16}
{Berg} T.~A.~M. {et~al.}, 2016, \mnras, 463, 3021

\bibitem[{{Berg} {et~al}\mbox{.}(2021){Berg}, {Fumagalli}, {D'Odorico},
  {Ellison}, {L{\'o}pez}, {Becker}, {Christensen}, {Cupani}, {Denney},
  {S{\'a}nchez-Ram{\'\i}rez}, \& {Worseck}}]{Berg21}
{Berg} T. A.~M. {et~al.}, 2021, \mnras, 502, 4009

\bibitem[{{Bond}(1980)}]{Bond1980}
{Bond} H.~E., 1980, \apjs, 44, 517

\bibitem[{{Bonifacio} {et~al}\mbox{.}(2009){Bonifacio}, {Spite}, {Cayrel},
  {Hill}, {Spite}, {Fran{\c{c}}ois}, {Plez}, {Ludwig}, {Caffau}, {Molaro},
  {Depagne}, {Andersen}, {Barbuy}, {Beers}, {Nordstr{\"o}m}, \&
  {Primas}}]{Bonifacio2009}
{Bonifacio} P. {et~al.}, 2009, \aap, 501, 519

\bibitem[{{Cayrel} {et~al}\mbox{.}(2004){Cayrel}, {Depagne}, {Spite}, {Hill},
  {Spite}, {Fran{\c{c}}ois}, {Plez}, {Beers}, {Primas}, {Andersen}, {Barbuy},
  {Bonifacio}, {Molaro}, \& {Nordstr{\"o}m}}]{Cayrel2004}
{Cayrel} R. {et~al.}, 2004, \aap, 416, 1117

\bibitem[{{Christensen} {et~al}\mbox{.}(2023){Christensen}, {Jakobsen},
  {Willott}, {Arribas}, {Bunker}, {Charlot}, {Maiolino}, {Marshall}, {Perna},
  \& {{\"U}bler}}]{Christensen23}
{Christensen} L. {et~al.}, 2023, \aap, 680, A82

\bibitem[{{Christlieb} {et~al}\mbox{.}(2001){Christlieb}, {Green}, {Wisotzki},
  \& {Reimers}}]{Christlieb2001}
{Christlieb} N., {Green} P.~J., {Wisotzki} L., {Reimers} D., 2001, \aap, 375,
  366

\bibitem[{{Clark} {et~al}\mbox{.}(2008){Clark}, {Glover}, \&
  {Klessen}}]{Clark08}
{Clark} P.~C., {Glover} S. C.~O., {Klessen} R.~S., 2008, \apj, 672, 757

\bibitem[{{Cooke} {et~al}\mbox{.}(2011){Cooke}, {Pettini}, {Steidel}, {Rudie},
  \& {Nissen}}]{Cooke11}
{Cooke} R., {Pettini} M., {Steidel} C.~C., {Rudie} G.~C., {Nissen} P.~E., 2011,
  \mnras, 417, 1534

\bibitem[{{Cooke} \& {Madau}(2014)}]{CookeMadau2014}
{Cooke} R.~J., {Madau} P., 2014, \apj, 791, 116

\bibitem[{{Cooke} {et~al}\mbox{.}(2015){Cooke}, {Pettini}, \&
  {Jorgenson}}]{Cooke15}
{Cooke} R.~J., {Pettini} M., {Jorgenson} R.~A., 2015, \apj, 800, 12

\bibitem[{{Cooke} {et~al}\mbox{.}(2014){Cooke}, {Pettini}, {Jorgenson},
  {Murphy}, \& {Steidel}}]{Cooke14}
{Cooke} R.~J., {Pettini} M., {Jorgenson} R.~A., {Murphy} M.~T., {Steidel}
  C.~C., 2014, \apj, 781, 31

\bibitem[{{Cooke} {et~al}\mbox{.}(2017){Cooke}, {Pettini}, \&
  {Steidel}}]{Cooke17}
{Cooke} R.~J., {Pettini} M., {Steidel} C.~C., 2017, \mnras, 467, 802

\bibitem[{{De Cia} {et~al}\mbox{.}(2018){De Cia}, {Ledoux}, {Petitjean}, \&
  {Savaglio}}]{deCia18}
{De Cia} A., {Ledoux} C., {Petitjean} P., {Savaglio} S., 2018, \aap, 611, A76

\bibitem[{{Dekker} {et~al}\mbox{.}(2000){Dekker}, {D'Odorico}, {Kaufer},
  {Delabre}, \& {Kotzlowski}}]{Dekker00}
{Dekker} H., {D'Odorico} S., {Kaufer} A., {Delabre} B., {Kotzlowski} H., 2000,
  in Society of Photo-Optical Instrumentation Engineers (SPIE) Conference
  Series, Vol. 4008, Optical and IR Telescope Instrumentation and Detectors,
  {Iye} M., {Moorwood} A.~F., eds., pp. 534--545

\bibitem[{{Dessauges-Zavadsky} {et~al}\mbox{.}(2009){Dessauges-Zavadsky},
  {Ellison}, \& {Murphy}}]{DessaugesZavadsky09}
{Dessauges-Zavadsky} M., {Ellison} S.~L., {Murphy} M.~T., 2009, \mnras, 396,
  L61

\bibitem[{{Dessauges-Zavadsky} {et~al}\mbox{.}(2003){Dessauges-Zavadsky},
  {P{\'e}roux}, {Kim}, {D'Odorico}, \& {McMahon}}]{DessaugesZavadsky03}
{Dessauges-Zavadsky} M., {P{\'e}roux} C., {Kim} T.~S., {D'Odorico} S.,
  {McMahon} R.~G., 2003, \mnras, 345, 447

\bibitem[{{D'Odorico} {et~al}\mbox{.}(2024){D'Odorico}, {Bolton},
  {Christensen}, {De Cia}, {Zackrisson}, {Kordt}, {Izzo}, {Li}, {Maiolino},
  {Marconi}, {Richter}, {Saccardi}, {Salvadori}, {Vanni}, {Feruglio},
  {Fumagalli}, {Fynbo}, {Noterdaeme}, {Papaderos}, {P{\'e}roux}, {Verma}, {Di
  Marcantonio}, {Origlia}, \& {Zanutta}}]{Dodorico23}
{D'Odorico} V. {et~al.}, 2024, Experimental Astronomy, 58, 21

\bibitem[{{Ellison} {et~al}\mbox{.}(2010){Ellison}, {Prochaska}, {Hennawi},
  {Lopez}, {Usher}, {Wolfe}, {Russell}, \& {Benn}}]{Ellison10}
{Ellison} S.~L., {Prochaska} J.~X., {Hennawi} J., {Lopez} S., {Usher} C.,
  {Wolfe} A.~M., {Russell} D.~M., {Benn} C.~R., 2010, \mnras, 406, 1435

\bibitem[{{Erni} {et~al}\mbox{.}(2006){Erni}, {Richter}, {Ledoux}, \&
  {Petitjean}}]{Erni06}
{Erni} P., {Richter} P., {Ledoux} C., {Petitjean} P., 2006, \aap, 451, 19

\bibitem[{{Ferland} {et~al}\mbox{.}(2017){Ferland}, {Chatzikos}, {Guzm{\'a}n},
  {Lykins}, {van Hoof}, {Williams}, {Abel}, {Badnell}, {Keenan}, {Porter}, \&
  {Stancil}}]{Cloudy17}
{Ferland} G.~J. {et~al.}, 2017, \rmxaa, 53, 385

\bibitem[{{Foreman-Mackey} {et~al}\mbox{.}(2013){Foreman-Mackey}, {Hogg},
  {Lang}, \& {Goodman}}]{emcee}
{Foreman-Mackey} D., {Hogg} D.~W., {Lang} D., {Goodman} J., 2013, \pasp, 125,
  306

\bibitem[{{Frebel} \& {Norris}(2015)}]{FrebelNorris2015}
{Frebel} A., {Norris} J.~E., 2015, \araa, 53, 631

\bibitem[{{Freudling} {et~al}\mbox{.}(2013){Freudling}, {Romaniello},
  {Bramich}, {Ballester}, {Forchi}, {Garc{\'{\i}}a-Dabl{\'o}}, {Moehler}, \&
  {Neeser}}]{ESOreflex}
{Freudling} W., {Romaniello} M., {Bramich} D.~M., {Ballester} P., {Forchi} V.,
  {Garc{\'{\i}}a-Dabl{\'o}} C.~E., {Moehler} S., {Neeser} M.~J., 2013, \aap,
  559, A96

\bibitem[{{Fumagalli} {et~al}\mbox{.}(2016{\natexlab{a}}){Fumagalli},
  {Cantalupo}, {Dekel}, {Morris}, {O'Meara}, {Prochaska}, \&
  {Theuns}}]{Fumagalli16a}
{Fumagalli} M., {Cantalupo} S., {Dekel} A., {Morris} S.~L., {O'Meara} J.~M.,
  {Prochaska} J.~X., {Theuns} T., 2016{\natexlab{a}}, \mnras, 462, 1978

\bibitem[{{Fumagalli} {et~al}\mbox{.}(2016{\natexlab{b}}){Fumagalli},
  {O'Meara}, \& {Prochaska}}]{Fumagalli16b}
{Fumagalli} M., {O'Meara} J.~M., {Prochaska} J.~X., 2016{\natexlab{b}}, \mnras,
  455, 4100

\bibitem[{{Glover}(2013)}]{Glover2013}
{Glover} S., 2013, in Astrophysics and Space Science Library, Vol. 396, The
  First Galaxies, {Wiklind} T., {Mobasher} B., {Bromm} V., eds., p. 103

\bibitem[{{Haardt} \& {Madau}(2012)}]{Haardt12}
{Haardt} F., {Madau} P., 2012, \apj, 746, 125

\bibitem[{{Hartwig} {et~al}\mbox{.}(2023){Hartwig}, {Ishigaki}, {Kobayashi},
  {Tominaga}, \& {Nomoto}}]{Hartwig23}
{Hartwig} T., {Ishigaki} M.~N., {Kobayashi} C., {Tominaga} N., {Nomoto} K.,
  2023, \apj, 946, 20

\bibitem[{{Hartwig} {et~al}\mbox{.}(2018){Hartwig}, {Yoshida}, {Magg},
  {Frebel}, {Glover}, {G{\'o}mez}, {Griffen}, {Ishigaki}, {Ji}, {Klessen},
  {O'Shea}, \& {Tominaga}}]{Hartwig18}
{Hartwig} T. {et~al.}, 2018, \mnras, 478, 1795

\bibitem[{{Heger} {et~al}\mbox{.}(2003){Heger}, {Fryer}, {Woosley}, {Langer},
  \& {Hartmann}}]{Heger03}
{Heger} A., {Fryer} C.~L., {Woosley} S.~E., {Langer} N., {Hartmann} D.~H.,
  2003, \apj, 591, 288

\bibitem[{{Heger} \& {Woosley}(2002)}]{Heger02}
{Heger} A., {Woosley} S.~E., 2002, \apj, 567, 532

\bibitem[{{Heger} \& {Woosley}(2010)}]{HegerWoosley2010}
{Heger} A., {Woosley} S.~E., 2010, \apj, 724, 341

\bibitem[{{Ishigaki} {et~al}\mbox{.}(2014){Ishigaki}, {Tominaga}, {Kobayashi},
  \& {Nomoto}}]{Ishigaki2014}
{Ishigaki} M.~N., {Tominaga} N., {Kobayashi} C., {Nomoto} K., 2014, \apjl, 792,
  L32

\bibitem[{{Ishigaki} {et~al}\mbox{.}(2018){Ishigaki}, {Tominaga}, {Kobayashi},
  \& {Nomoto}}]{Ishigaki2018}
{Ishigaki} M.~N., {Tominaga} N., {Kobayashi} C., {Nomoto} K., 2018, \apj, 857,
  46

\bibitem[{{Jeon} {et~al}\mbox{.}(2021){Jeon}, {Bromm}, {Besla}, {Yoon}, \&
  {Choi}}]{Jeon21}
{Jeon} M., {Bromm} V., {Besla} G., {Yoon} J., {Choi} Y., 2021, \mnras, 502, 1

\bibitem[{{Jofr{\'e}} {et~al}\mbox{.}(2015){Jofr{\'e}}, {Heiter}, {Soubiran},
  {Blanco-Cuaresma}, {Masseron}, {Nordlander}, {Chemin}, {Worley}, {Van Eck},
  {Hourihane}, {Gilmore}, {Adibekyan}, {Bergemann}, {Cantat-Gaudin},
  {Delgado-Mena}, {Gonz{\'a}lez Hern{\'a}ndez}, {Guiglion}, {Lardo}, {de
  Laverny}, {Lind}, {Magrini}, {Mikolaitis}, {Montes}, {Pancino},
  {Recio-Blanco}, {Sordo}, {Sousa}, {Tabernero}, \& {Vallenari}}]{Jofre15}
{Jofr{\'e}} P. {et~al.}, 2015, \aap, 582, A81

\bibitem[{{Kislitsyn} {et~al}\mbox{.}(2024){Kislitsyn}, {Balashev}, {Murphy},
  {Ledoux}, {Noterdaeme}, \& {Ivanchik}}]{Kislitsyn24}
{Kislitsyn} P.~A., {Balashev} S.~A., {Murphy} M.~T., {Ledoux} C., {Noterdaeme}
  P., {Ivanchik} A.~V., 2024, \mnras, 528, 4068

\bibitem[{{Klessen}(2019)}]{Klessen2019}
{Klessen} R., 2019, in Formation of the First Black Holes, {Latif} M.,
  {Schleicher} D., eds., pp. 67--97

\bibitem[{{Klessen} \& {Glover}(2023)}]{Klessen23}
{Klessen} R.~S., {Glover} S. C.~O., 2023, \araa, 61, 65

\bibitem[{{Ledoux} {et~al}\mbox{.}(2006){Ledoux}, {Petitjean}, {Fynbo},
  {M{\o}ller}, \& {Srianand}}]{Ledoux06}
{Ledoux} C., {Petitjean} P., {Fynbo} J.~P.~U., {M{\o}ller} P., {Srianand} R.,
  2006, \aap, 457, 71

\bibitem[{{Lind} \& {Amarsi}(2024)}]{Lind24}
{Lind} K., {Amarsi} A.~M., 2024, \araa, 62, 475

\bibitem[{{L{\'o}pez} {et~al}\mbox{.}(2016){L{\'o}pez}, {D'Odorico}, {Ellison},
  {Becker}, {Christensen}, {Cupani}, {Denney}, {P{\^a}ris}, {Worseck}, {Berg},
  {Cristiani}, {Dessauges-Zavadsky}, {Haehnelt}, {Hamann}, {Hennawi}, {Ir{\v
  s}i{\v c}}, {Kim}, {L{\'o}pez}, {Lund Saust}, {M{\'e}nard}, {Perrotta},
  {Prochaska}, {S{\'a}nchez-Ram{\'{\i}}rez}, {Vestergaard}, {Viel}, \&
  {Wisotzki}}]{Lopez16}
{L{\'o}pez} S. {et~al.}, 2016, \aap, 594, A91

\bibitem[{{Lopez} \& {Ellison}(2003)}]{Lopez03}
{Lopez} S., {Ellison} S.~L., 2003, \aap, 403, 573

\bibitem[{{Marassi} {et~al}\mbox{.}(2014){Marassi}, {Chiaki}, {Schneider},
  {Limongi}, {Omukai}, {Nozawa}, {Chieffi}, \& {Yoshida}}]{Marassi2014}
{Marassi} S., {Chiaki} G., {Schneider} R., {Limongi} M., {Omukai} K., {Nozawa}
  T., {Chieffi} A., {Yoshida} N., 2014, \apj, 794, 100

\bibitem[{{Milutinovic} {et~al}\mbox{.}(2010){Milutinovic}, {Ellison},
  {Prochaska}, \& {Tumlinson}}]{Milutinovic10}
{Milutinovic} N., {Ellison} S.~L., {Prochaska} J.~X., {Tumlinson} J., 2010,
  \mnras, 408, 2071

\bibitem[{{M{\o}ller} {et~al}\mbox{.}(2013){M{\o}ller}, {Fynbo}, {Ledoux}, \&
  {Nilsson}}]{Moeller13}
{M{\o}ller} P., {Fynbo} J.~P.~U., {Ledoux} C., {Nilsson} K.~K., 2013, \mnras,
  430, 2680

\bibitem[{{Morrison} {et~al}\mbox{.}(2016){Morrison}, {Kulkarni}, {Som},
  {DeMarcy}, {Quiret}, \& {P{\'e}roux}}]{Morrison16}
{Morrison} S., {Kulkarni} V.~P., {Som} D., {DeMarcy} B., {Quiret} S.,
  {P{\'e}roux} C., 2016, \apj, 830, 158

\bibitem[{{Murphy}(2018)}]{UVESpopler}
{Murphy} M., 2018, {Mtmurphy77/Uves\_Popler: Uves\_Popler: Post-Pipeline
  Echelle Reduction Software}

\bibitem[{{Neeleman} {et~al}\mbox{.}(2013){Neeleman}, {Wolfe}, {Prochaska}, \&
  {Rafelski}}]{Neeleman13}
{Neeleman} M., {Wolfe} A.~M., {Prochaska} J.~X., {Rafelski} M., 2013, \apj,
  769, 54

\bibitem[{{Nordlander} \& {Lind}(2017)}]{Nordlander17}
{Nordlander} T., {Lind} K., 2017, \aap, 607, A75

\bibitem[{{Noterdaeme} {et~al}\mbox{.}(2012){Noterdaeme}, {Petitjean},
  {Carithers}, {P{\^a}ris}, {Font-Ribera}, {Bailey}, {Aubourg}, {Bizyaev},
  {Ebelke}, {Finley}, {Ge}, {Malanushenko}, {Malanushenko},
  {Miralda-Escud{\'e}}, {Myers}, {Oravetz}, {Pan}, {Pieri}, {Ross},
  {Schneider}, {Simmons}, \& {York}}]{Noterdaeme12}
{Noterdaeme} P. {et~al.}, 2012, \aap, 547, L1

\bibitem[{{Penprase} {et~al}\mbox{.}(2010){Penprase}, {Prochaska}, {Sargent},
  {Toro-Martinez}, \& {Beeler}}]{Penprase10}
{Penprase} B.~E., {Prochaska} J.~X., {Sargent} W. L.~W., {Toro-Martinez} I.,
  {Beeler} D.~J., 2010, \apj, 721, 1

\bibitem[{{Pettini} {et~al}\mbox{.}(1994){Pettini}, {Smith}, {Hunstead}, \&
  {King}}]{Pettini94}
{Pettini} M., {Smith} L.~J., {Hunstead} R.~W., {King} D.~L., 1994, \apj, 426,
  79

\bibitem[{{Pettini} {et~al}\mbox{.}(2008){Pettini}, {Zych}, {Steidel}, \&
  {Chaffee}}]{Pettini08}
{Pettini} M., {Zych} B.~J., {Steidel} C.~C., {Chaffee} F.~H., 2008, \mnras,
  385, 2011

\bibitem[{{Poudel} {et~al}\mbox{.}(2020){Poudel}, {Kulkarni}, {Cashman},
  {Frye}, {P{\'e}roux}, {Rahmani}, \& {Quiret}}]{Poudel20}
{Poudel} S., {Kulkarni} V.~P., {Cashman} F.~H., {Frye} B., {P{\'e}roux} C.,
  {Rahmani} H., {Quiret} S., 2020, \mnras, 491, 1008

\bibitem[{{Prakapavi{\v{c}}ius} {et~al}\mbox{.}(2017){Prakapavi{\v{c}}ius},
  {Ku{\v{c}}inskas}, {Dobrovolskas}, {Klevas}, {Steffen}, {Bonifacio},
  {Ludwig}, \& {Spite}}]{Prakapavicius17}
{Prakapavi{\v{c}}ius} D., {Ku{\v{c}}inskas} A., {Dobrovolskas} V., {Klevas} J.,
  {Steffen} M., {Bonifacio} P., {Ludwig} H.~G., {Spite} M., 2017, \aap, 599,
  A128

\bibitem[{{Prochaska} \& {Wolfe}(1997)}]{Prochaska97}
{Prochaska} J.~X., {Wolfe} A.~M., 1997, \apj, 487, 73

\bibitem[{{Prochaska} \& {Wolfe}(2002)}]{Prochaska02}
{Prochaska} J.~X., {Wolfe} A.~M., 2002, \apj, 566, 68

\bibitem[{{Prochaska} \& {Wolfe}(2009)}]{Prochaska09}
{Prochaska} J.~X., {Wolfe} A.~M., 2009, \apj, 696, 1543

\bibitem[{{Quiret} {et~al}\mbox{.}(2016){Quiret}, {P{\'e}roux}, {Zafar},
  {Kulkarni}, {Jenkins}, {Milliard}, {Rahmani}, {Popping}, {Rao}, {Turnshek},
  \& {Monier}}]{Quiret16}
{Quiret} S. {et~al.}, 2016, \mnras, 458, 4074

\bibitem[{{Rafelski} {et~al}\mbox{.}(2012){Rafelski}, {Wolfe}, {Prochaska},
  {Neeleman}, \& {Mendez}}]{Rafelski12}
{Rafelski} M., {Wolfe} A.~M., {Prochaska} J.~X., {Neeleman} M., {Mendez} A.~J.,
  2012, \apj, 755, 89

\bibitem[{{Rix} {et~al}\mbox{.}(2007){Rix}, {Pettini}, {Steidel}, {Reddy},
  {Adelberger}, {Erb}, \& {Shapley}}]{Rix07}
{Rix} S.~A., {Pettini} M., {Steidel} C.~C., {Reddy} N.~A., {Adelberger} K.~L.,
  {Erb} D.~K., {Shapley} A.~E., 2007, \apj, 670, 15

\bibitem[{{Roederer} {et~al}\mbox{.}(2014){Roederer}, {Preston}, {Thompson},
  {Shectman}, {Sneden}, {Burley}, \& {Kelson}}]{Roederer2014}
{Roederer} I.~U., {Preston} G.~W., {Thompson} I.~B., {Shectman} S.~A., {Sneden}
  C., {Burley} G.~S., {Kelson} D.~D., 2014, \aj, 147, 136

\bibitem[{{Saccardi} {et~al}\mbox{.}(2023){Saccardi}, {Salvadori}, {D'Odorico},
  {Cupani}, {Fumagalli}, {Berg}, {Becker}, {Ellison}, \& {Lopez}}]{Saccardi23}
{Saccardi} A. {et~al.}, 2023, \apj, 948, 35

\bibitem[{{S{\'a}nchez-Ram{\'\i}rez}
  {et~al}\mbox{.}(2016){S{\'a}nchez-Ram{\'\i}rez}, {Ellison}, {Prochaska},
  {Berg}, {L{\'o}pez}, {D'Odorico}, {Becker}, {Christensen}, {Cupani},
  {Denney}, {P{\^a}ris}, {Worseck}, \& {Gorosabel}}]{SanchezRamirez16}
{S{\'a}nchez-Ram{\'\i}rez} R. {et~al.}, 2016, \mnras, 456, 4488

\bibitem[{{Sk{\'u}lad{\'o}ttir} {et~al}\mbox{.}(2018){Sk{\'u}lad{\'o}ttir},
  {Salvadori}, {Pettini}, {Tolstoy}, \& {Hill}}]{Skuladottir18}
{Sk{\'u}lad{\'o}ttir} {\'A}., {Salvadori} S., {Pettini} M., {Tolstoy} E.,
  {Hill} V., 2018, \aap, 615, A137

\bibitem[{{Sodini} {et~al}\mbox{.}(2024){Sodini}, {D'Odorico}, {Salvadori},
  {Vanni}, {Bischetti}, {Cupani}, {Davies}, {Becker}, {Ba{\~n}ados}, {Bosman},
  {Davies}, {Paolo Farina}, {Ferrara}, {Keating}, {Kulkarni}, {Lai},
  {Ryan-Weber}, {Maria Sebastian}, \& {Walter}}]{Sodini24}
{Sodini} A. {et~al.}, 2024, \aap, 687, A314

\bibitem[{{Stacy} {et~al}\mbox{.}(2010){Stacy}, {Greif}, \& {Bromm}}]{Stacy10}
{Stacy} A., {Greif} T.~H., {Bromm} V., 2010, \mnras, 403, 45

\bibitem[{{Starkenburg} {et~al}\mbox{.}(2017){Starkenburg}, {Martin},
  {Youakim}, {Aguado}, {Allende Prieto}, {Arentsen}, {Bernard}, {Bonifacio},
  {Caffau}, {Carlberg}, {C{\^o}t{\'e}}, {Fouesneau}, {Fran{\c{c}}ois},
  {Franke}, {Gonz{\'a}lez Hern{\'a}ndez}, {Gwyn}, {Hill}, {Ibata}, {Jablonka},
  {Longeard}, {McConnachie}, {Navarro}, {S{\'a}nchez-Janssen}, {Tolstoy}, \&
  {Venn}}]{Starkenburg2017}
{Starkenburg} E. {et~al.}, 2017, \mnras, 471, 2587

\bibitem[{{Sukhbold} {et~al}\mbox{.}(2016){Sukhbold}, {Ertl}, {Woosley},
  {Brown}, \& {Janka}}]{Sukhbold16}
{Sukhbold} T., {Ertl} T., {Woosley} S.~E., {Brown} J.~M., {Janka} H.~T., 2016,
  \apj, 821, 38

\bibitem[{{Susa} {et~al}\mbox{.}(2014){Susa}, {Hasegawa}, \&
  {Tominaga}}]{Susa14}
{Susa} H., {Hasegawa} K., {Tominaga} N., 2014, \apj, 792, 32

\bibitem[{{Tominaga} {et~al}\mbox{.}(2014){Tominaga}, {Iwamoto}, \&
  {Nomoto}}]{Tominaga2014}
{Tominaga} N., {Iwamoto} N., {Nomoto} K., 2014, \apj, 785, 98

\bibitem[{{Tumlinson} {et~al}\mbox{.}(2017){Tumlinson}, {Peeples}, \&
  {Werk}}]{Tumlinson17}
{Tumlinson} J., {Peeples} M.~S., {Werk} J.~K., 2017, \araa, 55, 389

\bibitem[{{Vanni} {et~al}\mbox{.}(2024){Vanni}, {Salvadori}, {D'Odorico},
  {Becker}, \& {Cupani}}]{Vanni24}
{Vanni} I., {Salvadori} S., {D'Odorico} V., {Becker} G.~D., {Cupani} G., 2024,
  \apjl, 967, L22

\bibitem[{{Vanni} {et~al}\mbox{.}(2023){Vanni}, {Salvadori},
  {Sk{\'u}lad{\'o}ttir}, {Rossi}, \& {Koutsouridou}}]{Vanni23}
{Vanni} I., {Salvadori} S., {Sk{\'u}lad{\'o}ttir} {\'A}., {Rossi} M.,
  {Koutsouridou} I., 2023, \mnras, 526, 2620

\bibitem[{{Vladilo} {et~al}\mbox{.}(2011){Vladilo}, {Abate}, {Yin}, {Cescutti},
  \& {Matteucci}}]{Vladilo11}
{Vladilo} G., {Abate} C., {Yin} J., {Cescutti} G., {Matteucci} F., 2011, \aap,
  530, A33

\bibitem[{{Vladilo} {et~al}\mbox{.}(2001){Vladilo}, {Centuri{\'o}n},
  {Bonifacio}, \& {Howk}}]{Vladilo01}
{Vladilo} G., {Centuri{\'o}n} M., {Bonifacio} P., {Howk} J.~C., 2001, \apj,
  557, 1007

\bibitem[{{Vladilo} {et~al}\mbox{.}(2018){Vladilo}, {Gioannini}, {Matteucci},
  \& {Palla}}]{Vladilo18}
{Vladilo} G., {Gioannini} L., {Matteucci} F., {Palla} M., 2018, \apj, 868, 127

\bibitem[{{Welsh} {et~al}\mbox{.}(2019){Welsh}, {Cooke}, \&
  {Fumagalli}}]{Welsh19}
{Welsh} L., {Cooke} R., {Fumagalli} M., 2019, \mnras, 487, 3363

\bibitem[{{Welsh} {et~al}\mbox{.}(2021){Welsh}, {Cooke}, \&
  {Fumagalli}}]{Welsh21}
{Welsh} L., {Cooke} R., {Fumagalli} M., 2021, \mnras, 500, 5214

\bibitem[{{Welsh} {et~al}\mbox{.}(2024){Welsh}, {Cooke}, {Fumagalli},
  {Pettini}, \& {Rudie}}]{Welsh24}
{Welsh} L., {Cooke} R., {Fumagalli} M., {Pettini} M., {Rudie} G.~C., 2024,
  \aap, 691, A285

\bibitem[{{Wolfe} {et~al}\mbox{.}(2005){Wolfe}, {Gawiser}, \&
  {Prochaska}}]{Wolfe05}
{Wolfe} A.~M., {Gawiser} E., {Prochaska} J.~X., 2005, \araa, 43, 861

\bibitem[{{Woosley}(2017)}]{Woosley17}
{Woosley} S.~E., 2017, \apj, 836, 244

\bibitem[{{Woosley} \& {Weaver}(1995)}]{Woosley95}
{Woosley} S.~E., {Weaver} T.~A., 1995, \apjs, 101, 181

\bibitem[{{Yong} {et~al}\mbox{.}(2013){Yong}, {Norris}, {Bessell},
  {Christlieb}, {Asplund}, {Beers}, {Barklem}, {Frebel}, \& {Ryan}}]{Yong2013}
{Yong} D. {et~al.}, 2013, \apj, 762, 26

\bibitem[{{Zafar} {et~al}\mbox{.}(2013){Zafar}, {P{\'e}roux}, {Popping},
  {Milliard}, {Deharveng}, \& {Frank}}]{Zafar13}
{Zafar} T., {P{\'e}roux} C., {Popping} A., {Milliard} B., {Deharveng} J.-M.,
  {Frank} S., 2013, \aap, 556, A141

\end{thebibliography}

\appendix

\section{Model parameter distributions}

\begin{figure*}
    \begin{center}
\includegraphics[width=\textwidth]{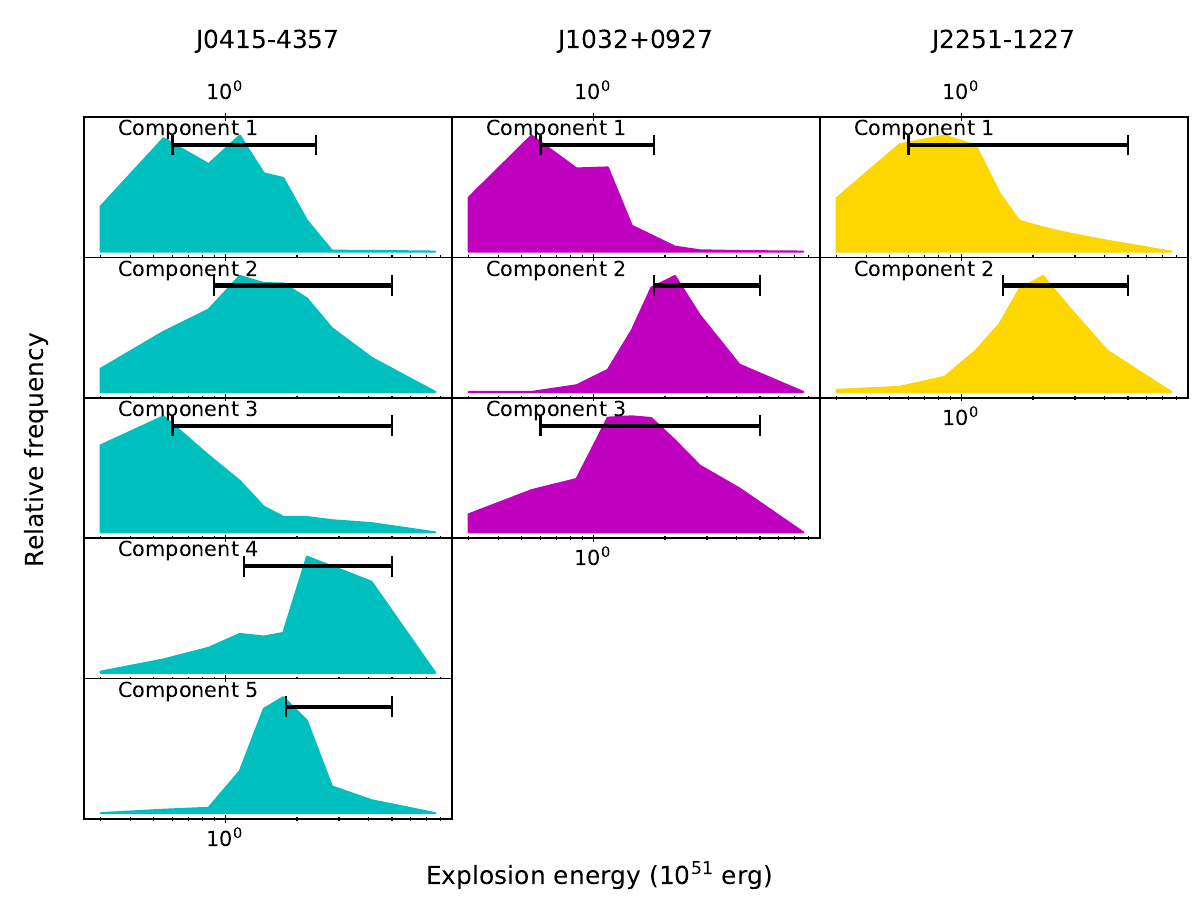}
\caption{The posterior distributions of the modelled core-collapse supernova explosion energy ($E_{\rm exp}$; in units of $10^{51}$ erg) for each component of the three subDLAs towards J0415$-$4357 (left column), J1032+0927 (middle), and J2251$-$1227 (right). The distributions have been smoothed using a top-hat kernel with a width of three bins. The notation is the same as in Figure \ref{fig:masshists}.}
\label{fig:exphists}
\end{center}
\end{figure*}

\begin{figure*}
    \begin{center}
\includegraphics[width=\textwidth]{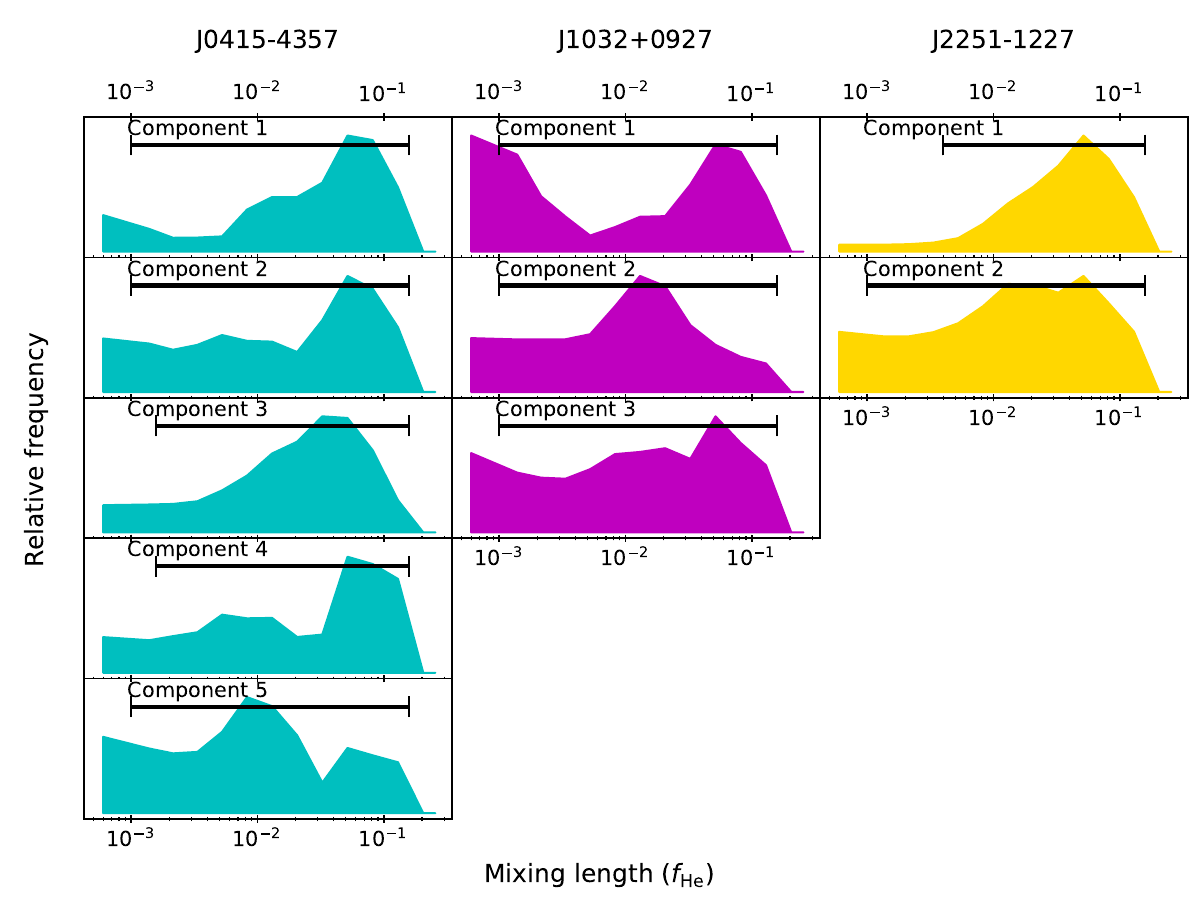}
\caption{The posterior distributions of the modelled mixing parameter ($f_{\rm He}$) for each component of the three subDLAs towards J0415$-$4357 (left column), J1032+0927 (middle), and J2251$-$1227 (right). The distributions have been smoothed using a top-hat kernel with a width of three bins. The notation is the same as in Figure \ref{fig:masshists}.}
\label{fig:mixhists}
\end{center}
\end{figure*}

\end{document}